\documentclass[prd,aps,a4paper,eqsecnum,nofootinbib,floatfix,twocolumn]{revtex4}  %
\newif\ifusesec
\usesectrue  
   
\usepackage{graphicx} 
\usepackage{amsmath,amsfonts,amssymb}
\usepackage{color}
\newcommand{\beq}{\begin{equation}}
\newcommand{\eeq}{\end{equation}}
\newcommand{\bea}{\begin{eqnarray}}
\newcommand{\eea}{\end{eqnarray}}
\newcommand{\g}{\gamma}
\newcommand{\pinf}{p_{\infty}}

\begin{document}

\title{High Precision Black Hole Scattering: Tutti Frutti vs Worldline Effective Field Theory}

\author{Donato Bini$^{1}$, Thibault Damour$^2$}
  \affiliation{
$^1$Istituto per le Applicazioni del Calcolo ``M. Picone,'' CNR, I-00185 Rome, Italy\\
$^2$Institut des Hautes Etudes Scientifiques, 91440 Bures-sur-Yvette, France
}

\date{\today}

\begin{abstract}
We consider black hole scattering up to the fifth Post Minkowskian ($G^5$) order and   
compare the predictions of the Tutti Frutti formalism to the results obtained within two different versions of  Worldline Effective Field Theory. At  the $G^4$ order we highlight  the complete agreement between Tutti Frutti results  and  the results of [C. Dlapa et al., Phys. Rev. Lett. \textbf{130}, no.10, 101401 (2023)], 
%\cite{Dlapa:2022lmu}
 and show how the Tutti Frutti approach allows one to extract the $O(G^3)$ angular momentum loss from the $O(G^4)$ impulse.
We compare the sixth Post-Newtonian (6PN) accurate Tutti Frutti predictions to the recent results of   [M. Driesse et al., arXiv:2411.11846 [hep-th]], % \cite{Driesse:2024feo}, 
which are  at the $G^5$ order, and at the leading order in the two mass  ratios, finding complete agreement.   We highlight that this agreement involves  the presence at the 5.5PN level of a nonlocal tail-of-tail contribution  to the scattering (first computed in  [D. Bini et al., Phys. Rev. D \textbf{102}, no.8, 084047 (2020)]),
%\cite{Bini:2020hmy}), 
and involves, at the 6PN level, the presence of a $O(G^4)$ contribution to the angular momentum loss [C. Heissenberg, arXiv:2501.02904 [hep-th]].
%\cite{Heissenberg:2025ocy}.
At the second order in the mass ratios of the $O(G^5)$ order we predict two independent  gauge-invariant observables to high-PN accuracy.
\end{abstract}

\maketitle

\section{Introduction}

Gravitational scattering has been the focus of an intense activity in recent years, both for conservative and dissipative (i.e. gravitational radiation related) effects. A number of approximation methods have been applied to this problem:
Post Newtonian (PN), see \cite{Blanchet:2013haa,Schafer:2018jfw}, 
Multipolar Post-Minkowskian (MPM) \cite{Blanchet:1985sp,Blanchet:1989ki,Damour:1990gj}, 
Non Relativistic General Relativity (NRGR) \cite{Goldberger:2004jt,Goldberger:2009qd,Foffa:2011ub,Porto:2016pyg,Foffa:2019hrb},
Gravitational Self Force (GSF) \cite{Barack:2018yvs},
Tutti Frutti (TF) \cite{Bini:2019nra,Bini:2020hmy,Bini:2020nsb,Bini:2020wpo,Bini:2020rzn}, and various avatars of 
Post Minkowskian (PM) perturbation theory, including Amplitude-based  Effective Field Theory approaches  and Worldline Effective Field Theory (WEFT), see, e.g., 
\cite{Amati:1990xe,Damour:2017zjx,Bjerrum-Bohr:2018xdl,Kosower:2018adc,Cheung:2018wkq,Bjerrum-Bohr:2019kec,
Bern:2019crd,Bern:2019nnu,Mogull:2020sak,Kalin:2020mvi,Kalin:2020fhe,
Bern:2021dqo,Bern:2021yeh,Dlapa:2021npj,Dlapa:2021vgp,Bjerrum-Bohr:2021din,
Saketh:2021sri,Kalin:2022hph,Khalil:2022ylj,Herrmann:2021tct,Mougiakakos:2021ckm,Jakobsen:2021smu,Riva:2021vnj,Manohar:2022dea,Dlapa:2022lmu,Driesse:2024xad}.

The aim of the present work is to relate several recent and ongoing WEFT  developments in the radiation reacted scattering of binary systems \cite{Dlapa:2022lmu,Driesse:2024feo} to state-of-the-art TF results \cite{Bini:2019nra,Bini:2020hmy,Bini:2020nsb,Bini:2020wpo,Bini:2020rzn}. 
Namely, we consider the radiatively corrected $O(G^4)$ scattering computations of Ref. \cite{Dlapa:2022lmu} and the recent $O(G^5)$ radiation reacted scattering results of Ref. \cite{Driesse:2024feo} and compare them to the results obtained in the TF approach \cite{Bini:2021gat,Bini:2022enm}.
The comparison between these results will be done at the level of the  observables $\Delta p_a^\mu$ (where $a=1,2$ is a body label and $\mu=0,1,2,3$ a spacetime index)
\bea
\label{DeltapaPM}
\Delta p_a^\mu&=&p_a^{{\rm (out)}\mu}-p_a^{{\rm (in)}\mu}\nonumber\\
&=&\Delta p_a^{G^1 \mu}+\Delta p_a^{G^2 \mu} +\ldots +\Delta p_a^{G^5 \mu}+O(G^6)\,,\qquad
\eea
measuring the change in four momentum of each body 
during scattering. $\Delta p_a^\mu$ is also known as the impulse of body $a$.
In the second line of Eq. \eqref{DeltapaPM} we have indicated the PM expansion of $\Delta p_a^\mu$, i.e., its expansion in powers of $G$ at fixed impact parameter $b$.
The initial four momentum $p_a^{{\rm (in)}\mu}$ of each body takes the form $p_a^{{\rm (in)}\mu}=m_a u_a^\mu$, where $m_a$ is the mass of body $a$ and  $u_a^\mu$ (satisfying $u_a\cdot u_a=-1$) its initial four velocity.
Each impulse admits a decomposition along the three 4-vectors:  $u_1^\mu$, $u_2^\mu$ and
\beq
\hat b^\mu \equiv \frac{b_{12}^\mu }{b}\equiv \frac{b_{1}^\mu-b_{2}^\mu  }{b}\,.
\eeq
The vectorial impact parameter $b_{12}^\mu $ is orthogonal to $u_1^\mu$ and $u_2^\mu$ and the scalar impact parameter $b$ denotes its magnitude (so $\hat b \cdot \hat b=1$ in our mostly plus signature). Denoting by $\gamma\equiv -u_1\cdot u_2$ the relative Lorentz factor of the two incoming worldlines, the decomposition of the impulse $\Delta p_a^\mu$ reads
\bea
\label{DeltapaPM1}
\Delta p_a^\mu&=&\overset{a}{C}_{b}\left[\frac{G}{b},\gamma, m_1,m_2\right] \hat b^\mu +\overset{a}{C}_{u_1}\left[\frac{G}{b},\gamma, m_1,m_2\right]u_1^\mu \nonumber\\
&+&\overset{a}{C}_{u_2}\left[\frac{G}{b},\gamma, m_1,m_2\right] u_2^\mu\,,
\eea 
or, equivalently, 
\bea
\label{DeltapaPM2}
\Delta p_a^\mu &=&\overset{a}{C}_{b}\left[\frac{G}{b},\gamma, m_1,m_2\right] \hat b^\mu +\overset{a}{C}_{\check u_1}\left[\frac{G}{b},\gamma, m_1,m_2\right]\check u_1^\mu\nonumber\\ 
&+&\overset{a}{C}_{\check u_2}\left[\frac{G}{b},\gamma, m_1,m_2\right] \check u_2^\mu\,,
\eea
where we introduced the dual four velocities $\check u_a^\mu$ defined as
\beq
\check u_1=\frac{u_1-\gamma u_2}{\gamma^2-1}\,,\quad \check u_2=\frac{u_2-\gamma u_1}{\gamma^2-1}\,,
\eeq
and such that $u_a \cdot \check u_b=+\delta_{ab}$ ($a,b=1,2$).
The various coefficients $\overset{a}{C}_{b}(\frac{G}{b},\gamma, m_1,m_2),...,\overset{a}{C}_{\check u_2}(\frac{G}{b},\gamma, m_1,m_2)$ admit an expansion in powers of $\frac{G}{b}$ (PM-expansion), and, at each order in $(\frac{G}{b})^n$ ($n\ge 1$), are polynomials of order $n+1$ in the two masses $m_1$ and $m_2$ of the form
\bea
\label{DeltapaPM3}
\overset{a}{C}{}^{G^n}_{X}(\gamma, m_1,m_2)&&\sim \left(\frac{G}{b}\right)^n m_1 m_2 \times \nonumber\\ 
&&(m_1^{n-1}+m_1^{n-2}m_2 +\dots + m_2^{n-1})\,.\qquad
\eea 
Each coefficient in the latter polynomial in masses is  only a function of $\gamma$. 
The PN expansion of the impulse at any given PM order $G^n$ is obtained by expanding the latter functions of $\gamma$ in a small velocity expansion around $\gamma=1$.
When combining the structures indicated in  Eqs. \eqref{DeltapaPM2} and \eqref{DeltapaPM3} the impulses read (where $\check k$ will be defined below)
\bea
\Delta p_a^\mu 
&=& m_1 m_2 \sum_{n=1}^\infty \sum_{k=0}^n G^n m_2^{n-k}m_1^k \Delta p_{a\,\check k{\rm SF}}^{(n)\,\mu}\nonumber\\
&=&
\overset{a}{C}_{b}\hat b^\mu +\overset{a}{C}_{\check u_1}\check u_1^\mu+\overset{a}{C}_{\check u_2} \check u_2^\mu\,.
\eea

In the following we will generally use as small velocity parameter the quantity
\beq
\label{pinfdef}
p_\infty=\sqrt{\gamma^2-1}=\frac{v}{\sqrt{1-v^2}}\,,
\eeq 
where $v$ is the relative velocity between the two incoming bodies \footnote{We mainly use units such that $c=1$.}.
By contrast to the WEFT computations which directly derive the total impulse in the above indicated polynomial form, the TF approach obtains the impulse in the form 
\beq
\Delta p_a^\mu  =\Delta p_a^{{\rm cons}\,\mu}  +\Delta p_a^{{\rm rrtot}\,\mu} \,, 
\eeq
where $\Delta p_a^{{\rm cons}\,\mu}$ is a conservative contribution (which starts at order $G^1$), while $\Delta p_a^{{\rm rrtot}\,\mu}$  denotes a radiation-reaction contribution (which starts at order $G^3$  and contains  squared of radiation reaction effects starting at order $G^4$).
In the TF formalism  both the conservative contribution  and the radiation reaction one are separately polynomial in masses at each order in $G$.

At the first two orders in $G$ we have (when considering body $1$)~\footnote{We denote the contribution of order $G^n$ to some quantity $Q$ either as $Q^{G^n}$ or $Q^{(n)}$.}
\bea
\label{deltap12}
\Delta p_1^{(1)\mu}&=& -\frac{G m_1 m_2}{b}2\chi_1 \hat b^\mu\nonumber\\
&=&-\frac{Gm_1 m_2}{b}\frac{2(2\gamma^2-1))}{\sqrt{\gamma^2-1}}\hat b^\mu\,,\nonumber\\
\Delta p_1^{(2)\mu}&=& -\frac{G^2m_1 m_2}{b^2}\left[2(m_1+m_2)\tilde \chi_2 \hat b^\mu\right.\nonumber\\
&-&\left. 2\chi_1^2 (m_1 \check u_2 -m_2\check u_1)  \right]\nonumber\\
&=&-\frac{G^2m_1 m_2}{b^2} \left[(m_1+m_2)\frac{3\pi (5\gamma^2-1)}{4\sqrt{\gamma^2-1}}\hat b^\mu \right.\nonumber\\
& -& \left. \frac{2(2\gamma^2-1)^2}{\gamma^2-1}
\left(m_1\check u_2^\mu -m_2\check u_1^\mu\right)\right]\,.
\eea
Here
\beq
\chi_1= \frac{2(2\gamma^2-1))}{\sqrt{\gamma^2-1}}\,,\quad \tilde \chi_2 =\frac{3\pi (5\gamma^2-1)}{4\sqrt{\gamma^2-1}}\,,
\eeq
are related to the PM expansion coefficients of the conservative scattering angles as
\beq
\chi=\sum_{n\ge 1} \frac{2\chi_n}{j^n}=h \sum_{n\ge 1}  2\tilde \chi_n \left(\frac{GM}{p_\infty b} \right)^n\,,
\eeq
where $M=m_1+m_2$,
\beq
j=\frac{J}{Gm_1m_2}=\frac{p_\infty b}{GM h}\,,\quad \tilde \chi_n=h^{n-1}\chi_n\,.
\eeq
Here
\beq
h=\frac{E}{M}=\sqrt{1+2\nu(\gamma-1)}\,,\quad \nu=\frac{m_1m_2}{(m_1+m_2)^2}\,,
\eeq
where $E$ denotes the (incoming) center-of-mass (c.m.) energy.
Note that $\tilde \chi_1\equiv \chi_1$. See Appendix \ref{app_notation} for a summary of our notation.

These impulses are conservative in the sense that they are symmetric (or better, equivariant) under time-reversal. Here,  $T$-symmetry can be technically defined 
by considering the velocity-parity properties of the ratio $\Delta p_1^\mu/p_\infty$. Indeed, under velocity-reversal such that $u_a^\mu \to -u_a^\mu$,    $p_\infty=\sqrt{\gamma^2-1}\to - p_\infty=-\sqrt{\gamma^2-1}$ (corresponding to $v\to -v$ in Eq. \eqref{pinfdef}), the ratios $\Delta p_1^{(1)\,\mu}/p_\infty$ and $\Delta p_1^{(2)\,\mu}/p_\infty$ are invariant. 

Starting at order $G^3$, where (using the notation of Ref. \cite{Driesse:2024feo})
%%%%
\bea
\label{deltap3}
\Delta p_1^{(3)\mu} &=&  m_1 m_2 (m_2^2 \Delta p^{(3)\mu}_{\rm 0SF}[\gamma, \hat b^\mu  ,\check u_1^\mu, \check u_2^\mu ]\nonumber\\ 
&+&  m_1 m_2  \Delta p^{(3)\mu}_{\rm 1SF}[\gamma, \hat b^\mu , \check u_1^\mu, \check u_2^\mu]\nonumber\\ 
&+&  m_1^2   \Delta p^{(3)\mu}_{\overline{\rm 0SF}}[\gamma, \hat b^\mu , \check u_1^\mu , \check u_2^\mu] )\,,
\eea
the dynamics is no longer $T$-symmetric, but can be uniquely decomposed in  a $T$-even,  conservative part (as measured by $\Delta p_1^{(3){\rm cons}\,\mu}/p_\infty$) and a $T$-odd radiation reaction part  (as measured by $\Delta p_1^{(3){\rm rr}\,\mu}/p_\infty$) with opposite time reversal properties. The TF approach decomposes $\Delta p_1^{(3)\mu}$ as follows, 
\bea
\label{deltap1G3}
\Delta p_1^{(3)\,\mu}&=& \Delta p_1^{(3){\rm cons}\,\mu}+\Delta p_1^{(3){\rm rr}\,\mu}\,,
\eea
where
\bea
\label{deltap1G3cons}
\Delta p_1^{(3){\rm cons}\,\mu} &=& -\frac{G^3 m_1 m_2(m_1+m_2)}{b^3}\left[\right.\nonumber\\
&&\frac{(m_1+m_2)}{\gamma^2-1}[2\tilde \chi_3^{\rm cons}-\frac43 h^2 (\chi_1^{\rm cons})^3]\hat b^\mu\nonumber\\ 
&+&\left. \frac{4\chi_1^{\rm cons}\tilde \chi_2^{\rm cons}}{\sqrt{\gamma^2-1}}(m_2\check u_1-m_1 \check u_2)  \right]\,,
\eea
and
\bea
\label{deltap1G3rr}
\Delta p_1^{(3){\rm rr}\mu} 
&=& \frac{G^3 m_1^2 m_2^2}{b^3}\left[-\frac{\tilde \chi_1^{\rm cons}\hat J_2(\gamma)}{(\gamma^2-1)}\hat b^\mu \right.\nonumber\\
&+&\left. \frac{\hat E_3(\gamma)}{(\gamma^2-1)^{3/2}}\check u_2  \right]\,.
\eea
Here $\hat J_2$ measures the $O(G^2)$ angular momentum loss \cite{Damour:2020tta,Bini:2022wrq}
\bea
\hat J_2(\gamma)&\equiv & h^2 J_2(\gamma,\nu) \equiv 2 (2\gamma^2-1)\sqrt{\gamma^2-1}{\mathcal I}(v)\,,\qquad
\eea
while  $\hat E_3$ measures the $O(G^3)$ energy loss \cite{Herrmann:2021lqe,Herrmann:2021tct}  
\bea
\hat E_3(\gamma)&\equiv &h^4E_3(\gamma,\nu)\equiv (\gamma^2-1)^{3/2} {\mathcal E}(\gamma)\,.
\eea
Here, $J_n(\gamma,\nu)$ and $E_n(\gamma,\nu)$ are defined so that
\bea
\label{JandEexp}
\frac{J_{\rm rad}}{J_{\rm c.m.}}&=& +\nu^1 \sum_{n=2}^\infty \frac{J_n}{j^n}\,,\nonumber\\
\frac{E_{\rm rad}}{m_1+m_2}&=& + \nu^2 \sum_{n=3}^\infty \frac{E_n}{j^n}\,.
\eea

At order $G^4$, where the impulse has the general mass structure
\bea
\label{eqp14}
\Delta p_1^{(4)\mu} &=&   m_1 m_2 (m_2^3 \Delta p^{(4)\mu}_{\rm 0SF}[\gamma, \hat b^\mu  ,\check u_1^\mu, \check u_2^\mu ]\nonumber\\ 
&+&  m_1 m_2^2 \Delta p^{(4)\mu}_{\rm 1SF}[\gamma, \hat b^\mu , \check u_1^\mu, \check u_2^\mu]\nonumber\\ 
&+&  m_1^2 m_2 \Delta p^{(4)\mu}_{\overline{\rm 1SF}} [\gamma,\hat b^\mu , \check u_1^\mu, \check u_2^\mu]\nonumber\\
&+&  m_1^3   \Delta p^{(4)\mu}_{\overline{\rm 0SF}}[\gamma, \hat b^\mu , \check u_1^\mu , \check u_2^\mu] )\,,
\eea
the (TF-defined) conservative part of the dynamics (as measured by $\Delta p_1^{(4){\rm cons}\,\mu}/p_\infty$, see the first of Eqs. \eqref{deltapamu2} below) is still $T$-even (and velocity-even), while the radiation reaction part is no longer purely $T$-odd but can be uniquely decomposed in a velocity-odd part (linear in radiation reaction) and a velocity-even part (quadratic in radiation reaction).
See detailed discussion in Section \ref{G4sect}.  

As will be discussed in Section \ref{G5sect} below, the situation at order $G^5$ becomes more involved because, though  the  (TF-defined) conservative part of the dynamics
is physically time-symmetric, the conservative impulse $\Delta p_1^{(4){\rm cons}\,\mu}/p_\infty$, starting at order 5.5 PN, also contains  odd powers of $p_\infty$. This progressive loss of time-reversal properties as the PM order increases is displayed in Table \ref{table_loss} where the words even and odd refer to the parity under $p_\infty \to - p_\infty$.

\begin{table}  
\caption{\label{table_loss} 
Progressive loss of velocity-reversal symmetry  properties as the PM order increases.
}
\begin{ruledtabular}
\begin{tabular}{lll}
$G^n$ &    cons  &  rr \\
\hline
$G^1$&    even  & -\\
\hline
$G^2$ &   even  & -\\
\hline
$G^3$ &    even  &   odd \\
\hline
$G^4$ &    even  &   odd + even  \\
\hline
$G^5$ &    even +  odd  &    odd + even \\
\end{tabular}
\end{ruledtabular}
\end{table}

A companion supplemental material file gives  explicit electronic forms of our main results (e.g., impulses at order $G^5$).

\section{Reminders of the structure of the impulse in the TF approach}

In the TF formalism the impulse is decomposed in a conservative part and a radiation reaction one (which includes terms linear and, starting at $G^4$, also  quadratic in radiation reaction):
\beq
\Delta p_a^\mu=\Delta p_a^{{\rm cons}\,\mu}+\Delta p_a^{{\rm rr\, tot}\,\mu}\,.
\eeq
 The conservative part of the impulse of body 1 reads 
\bea
\label{eq2.1}
\Delta p_1^{{\rm cons}\,\mu}  &=& 
-P_{\rm cm}\sin(\chi^{\rm cons})\hat b^\mu \nonumber\\ 
&+& (\cos(\chi^{\rm cons}) - 1)\frac{ m_1 m_2}{E^2} \left[(m_2+m_1\gamma ) u_1^\mu \right.\nonumber\\
&-&\left.  (m_1+m_2\gamma) u_2^\mu \right]\,.
\eea
We recall the general expression of the PM expansion of the (TF-defined) {\it conservative} angle $\chi^{\rm cons}$, namely
\bea
\label{chi_cons_def}
\chi^{\rm cons}&=&\sum_{n\ge 1} \frac{2\chi_n^{\rm cons}}{j^n}=\sum_{n\ge 1}\frac{2\tilde \chi^{\rm cons}_n(\gamma,\nu)}{h^{n-1}j^n}\nonumber\\
&=&h\sum_{n\ge 1}2\tilde \chi^{\rm cons}_n(\gamma,\nu)\left(\frac{G(m_1+m_2)}{p_\infty b }\right)^n\,,
\eea
 where $\tilde \chi^{\rm cons}_n(\gamma,\nu)=h^{n-1}\chi^{\rm cons}_n(\gamma,\nu)$ is a polynomial in $\nu$ of order $[\frac{n-1}{2}]$. Explicitly, $\tilde \chi^{\rm cons}_1\equiv \chi_1$ and $\tilde \chi^{\rm cons}_2$ are only functions of $\gamma$ while 
\bea
\label{eq_chi_tilde}
\tilde \chi^{\rm cons}_3(\gamma,\nu) &=& \tilde \chi^{0 \rm cons}_3(\gamma)+\nu \tilde \chi^{1 \rm cons}_3(\gamma)\,,\nonumber\\
\tilde \chi^{\rm cons}_4(\gamma,\nu) &=& \tilde \chi^{0 \rm cons}_4(\gamma)+\nu \tilde \chi^{1 \rm cons}_4(\gamma)\,,\nonumber\\
\tilde \chi^{\rm cons}_5(\gamma,\nu) &=& \tilde \chi^{0 \rm cons}_5(\gamma)+\nu \tilde \chi^{1 \rm cons}_5(\gamma)+ \nu^2 \tilde \chi^{2 \rm cons}_5(\gamma)\,.\nonumber\\
\eea
Here the $O(\nu^0)$ terms, namely $\tilde \chi^{\rm cons}_1$,  $\tilde \chi^{\rm cons}_2$ and all the $\tilde \chi^{0 \rm cons}_n(\gamma)$ ($n\geq 3$) are known from the probe limit of a test mass in a Schwarzschild background. See, e.g., Eqs. (8.1) of Ref. \cite{Bini:2020nsb}. For convenience they are reproduced in Table 
\ref{Schwchi} below.

The $\nu$-polynomiality structure of Eqs. \eqref{eq_chi_tilde} ensures that the corresponding impulse defined by Eq. \eqref{eq2.1} is a  polynomial in the two masses (as indicated in Eqs. 
\eqref{deltap12}, \eqref{deltap3}, \eqref{eqp14}).
More precisely, we can rewrite the conservative impulse as
\bea
\Delta p_1^{{\rm cons}\,\mu}&=& -2 P_{\rm c.m.}\sin \left( \frac{\chi^{\rm cons}}{2}\right) e_x^\mu\nonumber\\
&=& -2 m_1 m_2 p_\infty \frac{\sin \left( \frac{\chi^{\rm cons}}{2}\right)}{E}e_x^\mu\,,
\eea
with the unit vector 
\beq
e_x^\mu\equiv \hat b^{\mu}_{\rm eik}\equiv  \cos \frac{\chi_{\rm cons}}{2} \hat b^\mu  + \sin \frac{\chi_{\rm cons}}{2}   n^\mu\,,
\eeq
where
\beq
n^\mu =\frac{\sqrt{\gamma^2-1}}{E}(m_2 \check u_1^\mu -m_1 \check u_2^\mu)\,.  
\eeq
$e_x^\mu$ can be identified with the vectorial {\it eikonal} impact parameter used in Refs. \cite{Ciafaloni:2014esa,DiVecchia:2023frv}.  
 
In this decomposition both $\frac{\sin \left( \frac{\chi^{\rm cons}}{2}\right)}{E}$ and $e_x^\mu$ are polynomial in the masses (see Eq. (11.21) of Ref.  \cite{Bini:2022enm}), as shown in Appendix \ref{appendixex}.

The total radiative contribution to the impulses is obtained in the TF approach as a sum of various contributions, namely
\begin{widetext}
\beq
\label{finale}
\Delta p_{a\,\mu}^{\rm rr\, tot}=\chi^{\rm  rr \, rel } \frac{d}{d\chi_{\rm cons}}\Delta p_{a \mu}^{\rm cons} + 
\frac{\Delta P_{\rm c.m.}}{ P_{\rm c.m.}} p_{a \mu}^{\rm out} -\frac{m_a^2}{E_a} \frac{\Delta P_{\rm c.m.}}{ P_{\rm c.m.}} U_\mu-\frac{E_a}{E}   P_\mu^{\rm rad} - \frac{(p_{a \nu}^{\rm out}    P^\nu_{\rm rad})}{E} U_\mu+ \Delta  p_{a \, \mu}^{\rm rr \, remain }\,,
\eeq
\end{widetext}
where the first contribution is the usual linear-response contribution with \cite{Bini:2012ji} 
\beq
 \chi^{\rm  rr \, rel }\equiv - \left(\frac12 \frac{\partial \chi^{\rm cons}}{\partial  E  } E_{\rm rad} +\frac12 \frac{\partial \chi^{\rm cons}}{\partial  J } J_{\rm rad} \right) \,, 
\eeq
where the next four contributions are linked to recoil effects \cite{Bini:2021gat,Bini:2022enm}, 
\beq
\Delta P_{\rm c.m.} =  - \frac{E_1 E_2}{E P_{\rm c.m.}} E_{\rm rad}\,,
\eeq
and where $\Delta  p_{a }^{\rm rr \, remain }$ (which starts at order $G^4/c^{10}$) was introduced in Eq. (12.21) of Ref. \cite{Bini:2022enm}.
As shown there, the latter \lq\lq remain" contribution is constrained by three different  requirements: i) the antisymmetric constraint (12.23) of Ref. \cite{Bini:2022enm};
ii) its second-gravitational-self-force\footnote{As shown below, there is a distinction (as well as a relation) between the usual concept of 2GSF (linked to the square of radiation-reaction on a particle of mass $m_1$) and the notation 1SF/${\overline {\rm 1SF}}$/2SF introduced by \cite{Driesse:2024feo} and used already in Eq. \eqref{deltap3} abobve, and Eqs. \eqref{plefka_eqs} below.} (2GSF) character, i.e., the fact that it contains a factor $m_a^3$; and iii) the property of canceling the non polynomiality in masses of several terms entering Eq. \eqref{finale}. [Indeed, Eq. \eqref{finale} contains denominators such that $E_a$,  $P_{\rm c.m.}$, and $E$ which, considered by themselves,  violate the mass polynomiality of  $\Delta  p_{a }$.]  

It was shown in Ref. \cite{Bini:2022enm} that the requirements i), ii) iii) determine most of the structure of $\Delta  p_{1}^{ \rm rr \, remain }{}^\mu$ at orders $G^4$ and $G^5$, modulo a supplementary $O(G^5)$ term denoted   $\Delta  p_{1}^{ \rm rr \, remain \, sup}{}^\mu$ below. In other words, we have
\beq
\label{newdeco}
\Delta  p_{1}^{ \rm rr \, remain }{}^\mu=\Delta  p_{1}^{ {\rm rr \, remain}\, P_{\rm rad}}{}^\mu+\Delta  p_{1}^{ \rm rr \, remain \, sup}{}^\mu\,,
\eeq
where $\Delta p_1^{{\rm rr, remain} P_{\rm rad}^\mu}\sim G^4 +G^5 $ is fully determined by the TF approach, and is {\it linear} in $P_{\rm rad}^\mu$, and where
\beq
\label{eqremainsup}
\Delta  p_{1}^{ \rm rr \, remain \, sup}{}^\mu=\frac{G^5}{b^5}m_1^2 m_2^3 f_b^{G^5 \rm remain }\hat b^\mu +O(G^6)\,.
\eeq

In view of the decomposition  \eqref{newdeco} it is  convenient to gather all the  contributions linear in $P_{\rm rad}^\mu$,  and to define
\bea
\label{deltaplinP}
\Delta  p_{a }^{{\rm rr  lin} P_{\rm rad}}{}_\mu &\equiv & -  \frac12 \frac{\partial \chi^{\rm cons}}{\partial  E  } E_{\rm rad} \frac{d}{d\chi_{\rm cons}}\Delta p_{a \mu}^{\rm cons}\nonumber\\ 
&+& 
\frac{\Delta P_{\rm c.m.}}{ P_{\rm c.m.}} p_{a \mu}^{\rm out} -\frac{m_a^2}{E_a} \frac{\Delta P_{\rm c.m.}}{ P_{\rm c.m.}} U_\mu\nonumber\\
&-& \frac{E_a}{E}   P_\mu^{\rm rad} - \frac{(p_{a \nu}^{\rm out}    P^\nu_{\rm rad})}{E} U_\mu \nonumber\\
&+& \Delta  p_{a \mu}^{{\rm rr \, remain}\, P_{\rm rad} }\,.
\eea
The total impulse is then naturally decomposed in the TF formalism as follows:
\bea
\label{deco_fin}
\Delta p_a^\mu  &=& \Delta p_a^{{\rm cons}\,\mu}+\Delta  p_{a }^{{\rm rr  lin} J_{\rm rad}\mu}\nonumber\\
&+&\Delta  p_{a }^{{\rm rr  lin} P_{\rm rad}\mu}+\Delta  p_{1}^{ \rm rr \, remain \, sup}{}^\mu\,.
\eea
Here $\Delta  p_{a }^{{\rm rr  lin} J_{\rm rad} \mu}$ is the part of the linear response formula that is {\it linear} in $J_{\rm rad}$, namely \footnote{As discussed below, contributions nonlinear in $J_{\rm rad}$ start at order $G^6$.},  
\bea
\label{deltapradJ}
\Delta  p_{a }^{{\rm rr  lin} J_{\rm rad}}&\equiv & - \frac12 \frac{\partial \chi^{\rm cons}}{\partial  J } J_{\rm rad}  \frac{d}{d\chi_{\rm cons}}\Delta p_{a \mu}^{\rm cons} \,,
\eea
while $\Delta  p_{a }^{{\rm rr  lin} P_{\rm rad}\mu}$, defined in Eq. \eqref{deltaplinP}, 
is {\it linear} in the radiated four momentum $P_{\rm rad}^\mu$. We recall that the last term $\Delta  p_{1}^{ \rm rr \, remain \, sup}{}^\mu
$ in Eq. \eqref{deco_fin} starts at $G^5$ (see Eq. \eqref{eqremainsup}). The decomposition \eqref{deco_fin} generalizes the structure that was already present at $G^3$, see Eq.  \eqref{deltap1G3}, where the radiation reaction part given in Eq.  \eqref{deltap1G3rr} is linear in $\hat J_2$ and $\hat E_3$.

Each part  in the decomposition \eqref{deco_fin}  has the property of being polynomial in the two masses. 

The mass polynomiality of $\Delta p_a^{{\rm cons}\,\mu}$ follows from the $\nu-$structure of the conservative scattering angle displayed in Eq. \eqref{eq_chi_tilde} above.  The mass polynomiality of $\Delta  p_{a }^{{\rm rr  lin} J_{\rm rad}}$ follows from the $\nu-$structure of $J_{\rm rad}$ enunciated in Refs. \cite{Bini:2021gat,Bini:2022enm}, see Eq. (7.18) and (7.19)  in Ref. \cite{Bini:2021gat} and Eq. (10.3) in Ref.  \cite{Bini:2022enm} namely,
\bea
\hat J_n(\gamma,\nu)&\equiv & h^n J_n +h^{n-1}\nu E_n  \nonumber\\
&=& \hat J_n^0(\gamma) +\nu  \hat J_n^1(\gamma)+\ldots +  \hat J_n^{d(n)}(\gamma)\nu^{d(n)}\,,\qquad
\eea
where $d(n)$ is the integer part of $(n-2)/2$:
$
d(n)=\left[ \frac{n-2}{2}\right]\,,
$
and where the coefficients $E_n$ and $J_n$ denote the PM expansion coefficients of $J_{\rm rad}$ and $E_{\rm rad}$ defined in Eqs. \eqref{JandEexp}.
In particular,  the latter polynomiality  rule yields 
\bea
\label{hatJ_defs}
h^2 J_2&=&\hat J_2(\gamma)\,,\nonumber\\
h^3 J_3+h^2 \nu E_3&=&\hat J_3(\gamma)\,,\nonumber\\
h^4 J_4 +h^3 \nu E_4&=& \hat J_4^0(\gamma)+\nu  \hat J_4^1(\gamma)\,.
\eea
Finally, the mass polynomiality of $\Delta  p_{a }^{{\rm rr  lin} P_{\rm rad}}$ follows from the structure of $\Delta  p_{a }^{{\rm rr \, remain}\, P_{\rm rad} }$ together with the mass polynomiality structure of $P^\mu_{\rm rad}$, which also implies $\nu-$polynomiality rules of the $x$  and $y$ components of $P^\mu_{\rm rad}$, as discussed in Section X and XI of Ref. \cite{Bini:2022enm}. Let us, in particular, recall the $\nu-$polynomiality rule (which follows from Eq. (11.18) of Ref. \cite{Bini:2022enm})
\beq
\label{hatEn}
\hat E_n=h^{n+1}E_n=P_{[\frac{n-2}{2}]}(\nu)\,.
\eeq

\section{Comparison between the $O(G^4)$ TF impulse and WEFT}
\label{G4sect}

At order $G^4$, the three contributions (conservative, linear in $J_{\rm rad}$, linear in $P_{\rm rad}$) of the TF-predicted impulse, namely
\bea
\label{deltapamu1}
\Delta p_a^{G^4\,\mu}  &=&\left[\Delta p_a^{{\rm cons}\,\mu}+\Delta  p_{a }^{{\rm rr  lin} J_{\rm rad}}\right.\nonumber\\
&+&\left.\Delta  p_{a }^{{\rm rr  lin} P_{\rm rad}}\right]_{G^4}\,,
\eea
read
\bea
\label{deltapamu2}
\Delta p_1^{G^4 {\rm cons}\,\mu}&=& -2P_{\rm c.m.}\sin\frac{\chi^{\rm cons}}{2}e_x^\mu \bigg|_{G^4 }\,,\nonumber\\
\Delta  p_{1}^{G^4 {\rm rr  lin} J_{\rm rad}\, \mu}&=& \frac{G^4 m_1^2 m_2^2}{b^4}\left[ \frac{(m_1+m_2)\chi_1^{\rm cons}}{(\gamma^2-1)^{3/2}}\hat J_3 \hat b^\mu \right.\nonumber\\ 
&-&  2\frac{\hat J_2}{(\gamma^2-1)}[(\chi_1^{\rm cons})^2 (m_2\check u_1^\mu-m_1\check u_2^\mu)\nonumber\\
&+& \left.
\frac{(m_1+m_2)\tilde \chi_2^{\rm cons}}{\sqrt{\gamma^2-1}}\hat b^\mu]\right]\,,
\nonumber\\
\Delta  p_{1}^{G^4 {\rm rr  lin} P_{\rm rad}\, \mu}&=& \frac{G^4 m_1^2 m_2^2}{b^4}\left[m_1 \hat P_b^4 \hat b^\mu\right.\nonumber\\
&-&(\gamma-1)(m_1-m_2)\hat P^4_{1-2}\check u_2  \nonumber\\
&+& (\gamma+1)(m_1+m_2)\hat P^4_{1+2}\check u_2 \nonumber\\
&+&\frac{(m_1+m_2)}{\gamma-1}\hat P_{1+2}^3[2\gamma \chi_1\nonumber\\
&+&\left. (\gamma^2-1)\chi_1']\hat b^\mu\right]\,.\nonumber\\
\eea

Here, 
$\Delta p_1^{\mu}$ at order $G^4$, has been expressed in terms of $\hat J_2$, $\hat J_3$, and of the form factors $\hat P_{1+2}(\gamma)$, $\hat P_{1-2}(\gamma)$ and $\hat P_b(\gamma)$ of $P_\mu^{\rm rad}$ defined as
\beq
\label{Praddecom}
P_{\rm rad}^\mu=P_b \hat b^\mu +P_{1+2}(u_1^\mu+u_2^\mu)+ P_{1 - 2}(u_1^\mu-u_2^\mu)\,,
\eeq
with mass-factorized expressions at orders $G^3$ and $G^4$.
At order  $G^3$   only one function of $\gamma$ enters. Indeed, $P_b^{G^3}=0$, $P_{1-2}^{G^3}=0$ and
\beq
\label{P1p2G3}
P_{1+2}^{G^3}
=\frac{G^3 m_1^2 m_2^2 }{b^3} \hat P_{1+2}^3(\gamma)\,.
\eeq
At order  $G^4$ the various components of $P_{\rm rad}^\mu$ involve (after factoring the masses)  three functions of $\gamma$, namely 
\bea
\label{Pb1p21m2}
P_b^{G^4} 
&=& \frac{G^4 m_1^2 m_2^2 }{b^4}(m_2-m_1)\hat P_b^4(\gamma)\,,\nonumber\\
P_{1-2}^{G^4}
&=& \frac{G^4 m_1^2 m_2^2 }{b^4}(m_2-m_1)\hat P_{1-2}^4(\gamma)\,,\nonumber\\ 
P_{1+2}^{G^4}
&=& \frac{G^4 m_1^2 m_2^2 }{b^4}(m_1+m_2)   \hat P_{1+2}^4(\gamma)\,.
\eea
The form factor $\hat P_{1+2}^3(\gamma)$  is related to $\hat E_3=h^4 E_3=(\gamma^2-1)^{3/2}{\mathcal E}(\gamma)$ by
\beq
\hat P_{1+2}^3(\gamma)= \frac{{\mathcal E}(\gamma)}{\gamma+1}\,,
\eeq
while at order $G^4$ we have $\hat E_4=h^5 E_4=\tilde E_4^0+\nu \tilde E_4^1$, with the relations
\bea
\hat P_{1-2}^4(\gamma)&=&-\frac{1}{4(\gamma-1)}\frac{\tilde E_4^1}{(\gamma^2-1)^2}\,, \nonumber\\
\hat P_{1+2}^4(\gamma)&=&\frac{1}{4(\gamma+1)} \frac{4\tilde E_4^0+\tilde E_4^1}{(\gamma^2-1)^2}\,. 
\eea

\typeout{Put later: $\Delta p_a^{\rm cons}$ is time-even and TF has computed at 6PN and the result agrees with \cite{Bern:2021yeh,Dlapa:2021vgp,Bern:2022jvn}.}
It is convenient in the following to decompose the various contributions to $\Delta p_a$ according to the mass polynomiality as in Eq. \eqref{eqp14}.
Each $\Delta p_{\rm nSF}^\mu[\gamma, \hat b^\mu  ,\check u_1^\mu, \check u_2^\mu ]$ will be decomposed along $\hat b^\mu$, $\check u_1^\mu$, $\check u_2^\mu$ according to the following notation
\bea
\Delta p_{\rm nSF}^\mu[\gamma, \hat b^\mu  ,\check u_1^\mu, \check u_2^\mu ]&=&
 \hat b^\mu  f_b^{\rm nSF}(\gamma) + \check u_1^\mu  f_{\check u_1}^{\rm nSF}(\gamma)\nonumber\\ 
&+& \check u_2^\mu  f_{\check u_2}^{\rm nSF}(\gamma)\,,
\eea
where, the index $n$ takes the values   $n=0,1,\bar 1, \bar 0$ at $G^4$, and the values   $n=0,1,\bar 1,2, \bar 0$ at $G^5$. [Here, the notation $\bar 1$SF refers to  ${\overline {\rm 1SF}}$.]

The $G^4$ contributions to the various $f_X^{\rm nSF}(\gamma)$ (suppressing an overall factor $G^n/b^n$ to ease notation)
 predicted by the TF formalism have the following structure (in which the general notation $K$ refers to contributions   known  from lower PM  orders, e.g. in $f_b^{G^4 {\rm TF},{\rm 1SF}}$ $K$ stands for a function involving $\chi_1$, $\tilde \chi_2^{\rm cons}$, $\tilde \chi_4^{0\, {\rm cons}}$ and $\hat J_2$),
\bea
\label{G4sf1}
f_b^{G^4 {\rm TF},{\rm 1SF}}&=&-\frac{2}{(\gamma^2-1)^{3/2}}\tilde \chi_4^{1{\rm cons}}\nonumber\\ 
&-&\frac{1}{(\gamma^2-1)^{3/2}}\tilde \chi_1^{1{\rm cons}}\hat J_3 +K \,,\nonumber\\
f_{\check u_1}^{G^4 {\rm TF},{\rm 1SF}}&=& K \,, \nonumber\\
f_{\check u_2}^{G^4{\rm TF}, {\rm 1SF}}&=&(\gamma-1)\hat P_{1-2}^4+(\gamma+1)\hat P_{1+2}^4  +K\,,
\eea
while the ${\overline {\rm 1SF}}$ contributions read 
\bea
\label{G4sf2}
f_b^{G^4 {\rm TF},{\overline {\rm 1SF}}}&= & \hat P_b^4 -\frac{2}{b^4(\gamma^2-1)^{3/2}}\tilde \chi_4^{1{\rm cons}}\nonumber\\
& -&\frac{1}{(\gamma^2-1)^{3/2}}\tilde \chi_1^{1{\rm cons}}\hat J_3 +K\,, \nonumber\\
f_{\check u_1}^{G^4{\rm TF},{\overline {\rm 1SF}}}&= & K\,, \nonumber\\
f_{\check u_2}^{G^4{\rm TF},{\overline {\rm 1SF}}}&= &-(\gamma-1)\hat P_{1-2}^4+(\gamma+1)\hat P_{1+2}^4 +K \,.\qquad
\eea
One can recognize on the rhs of Eqs. \eqref{G4sf1}
 and \eqref{G4sf2} above the origin of the various contributions. For instance: 1) the terms $\propto \tilde \chi_4^{1{\rm cons}}$ come  from the conservative impulse $\Delta p_1^{\rm cons}$; 2) the terms $\hat J_3$ come from $\Delta  p_{}^{{\rm rr  lin} J_{\rm rad}}$, while 3) the terms $\propto \hat P^4_{1+ 2}$, $\propto \hat P^4_{1-2}$ or 
$\propto \hat P^4_b$ all come from $\Delta  p_{a }^{{\rm rr  lin} P_{\rm rad}}$.

At the $G^4$ order, the conservative contribution  to the impulse $\Delta p_a^{\rm cons}$  has the same meaning in TF and in WEFT. The exact value of 
$\Delta p_a^{G^4,\rm cons}$ has been computed in Refs. \cite{Bern:2021yeh,Dlapa:2021npj,Dlapa:2021vgp,Bern:2022jvn,Bjerrum-Bohr:2022ows}. The corresponding conservative TF impulse (which was computed earlier at the 6PN accuracy \cite{Bini:2021gat}) was found to be in full agreement with the WEFT results, up to the 6PN accuracy.

At this  $G^4$ order, the radiation reaction contribution  to the impulse $\Delta p_a^{G^4 \rm rr}$ can be defined both in TF and WEFT as being
\bea
\Delta p_a^{G^4 {\rm rr}\,\mu}&\equiv &\Delta p_a^{G^4\mu} -\Delta p_a^{G^4 {\rm cons}\,\mu}\,.
\eea
The exact  value of $\Delta p_a^{G^4 \rm rr}$ has been computed by Refs. \cite{Dlapa:2022lmu,Damgaard:2023ttc}.
  
Ref. \cite{Dlapa:2022lmu} used a decomposition of $\Delta p_a^{G^4 \rm rr}$ in terms of dissipative-1rad and dissipative-2rad contributions, which are defined by  their character under velocity-reversal: the 1rad part being $T$-odd and the 2rad part being $T$-even. A similar decomposition makes sense in the TF approach and was used in Ref. \cite{Bini:2022enm}. In particular, Eq. (12.34) of Ref.  \cite{Bini:2022enm} at the $G^4$ level has shown that the radiation-reaction part of the $b$-projection of $\Delta p_1$, namely $\Delta p_{1b}^{G^4 \rm rr}=\Delta p_{1bG^4} - \Delta p_{1bG^4}^{\rm cons}$ was given in the TF approach by
\bea
\label{deltap1brrG4}
\Delta p_{1b}^{G^4 \rm rr}
&=& \Delta p_{1bG^4}^{\rm rr, lin-odd}+\frac{m_1}{m_2-m_1}P_{xG^4}^{\rm rad}\,,
\eea
where $\Delta p_{1bG^4}^{\rm rr, lin-odd}$ is $T$-odd while $\frac{m_1}{m_2-m_1}P_{xG^4}^{\rm rad}$ is $T$-even. 
In terms of the decomposition in Eqs. \eqref{deltapamu1} and \eqref{deltapamu2},  the TF $T$-odd part is equal to
\beq
\Delta p_{1bG^4}^{\rm rr, lin-odd}=\hat b\cdot \Delta  p_{1}^{{\rm rr  lin} J_{\rm rad}}+\hat b \cdot \Delta  p_{1}^{{\rm rr  lin} P_{\rm rad}} |_{T-{\rm odd}}
\eeq
where the $T$-odd part of $\hat b \cdot \Delta  p_{1}^{{\rm rr  lin} P_{\rm rad}} $ is obtained from the third Eq. \eqref{deltapamu2} in the following two-step way.
First, one replaces $\hat P_b^4$
on the right-hand-side  of Eq. \eqref{deltapamu2} by
\beq
\label{Pb4vxPxy}
\hat P_b^4=-\frac{h^4}{(\gamma^2-1)^2}\left(-\check P_x^4+\chi_1^{\rm cons}\check P_y^3\right)\,,
\eeq
where $\check P_x^4$ and $\check P_y^3$ are related to the $1/j$ expansion of 
\beq
P_x\equiv P_{\rm rad}{}_\mu e_x^\mu\,,\quad P_y\equiv P_{\rm rad}{}_\mu e_y^\mu\,,
\eeq
in the following way
\bea
P_x&=&(m_2-m_1)\nu^2 \sum_{n\ge 4}\frac{\check P_x^n}{j^n}\,,\nonumber\\
P_y&=&(m_2-m_1)\nu^2 \sum_{n\ge 3}\frac{\check P_y^n}{j^n}\,.
\eea
We recall that the unit vector $e_x^\mu$ is defined as
\beq
e_x^\mu\equiv \hat b^{\mu}_{\rm eik}\equiv  \cos \frac{\chi_{\rm cons}}{2} \hat b^\mu  + \sin \frac{\chi_{\rm cons}}{2}   n^\mu\,,
\eeq 
see Eq. (3.49) in Ref. \cite{Bini:2021gat}. 
As a consequence $e_x^\mu$ has a nontrivial PM expansion so that 
\beq
e_x^\mu=\hat b^\mu + G e_x^{G^1\mu} +\ldots\,,
\eeq
which implies, in particular, that $[P_x]^{G^4}$ is the sum of a term  proportional to $P_b^{G^4}$ and a term linear in $P_{\rm rad}^{G^3}$ (as was already exhibited in Eq. \eqref{Pb4vxPxy}).

In addition, $\check P_y^3$ in Eq. \eqref{Pb4vxPxy} is given by
(see Eqs. (7.15) of Ref.\cite{Bini:2021gat}) 
\beq
\check P_y^3=\sqrt{\frac{\gamma-1}{\gamma+1}}E_3\,,\qquad h^4E_3=\hat E_3=(\gamma^2-1)^{3/2}{\mathcal E}(\gamma)\,.
\eeq
In the second step of the definition of the $T$-odd part of $\hat b \cdot \Delta  p_{1}^{{\rm rr  lin} P_{\rm rad}}$
one sets to zero the term proportional to $\check P_x^4$ (which is $T$-even, contrary to the $\check P_y^3$ term which is $T$-odd).

This leads us to the following  identifications between WEFT and TF quantities
\bea
\label{c1and2rad}
c^{4 \rm diss}_{\rm 1b,1rad}&=&\Delta p_{1bG^4}^{\rm rr, lin-odd}\,,\nonumber\\
c^{4 \rm diss}_{\rm 1b,2rad}&=&\frac{m_1}{m_2-m_1}P_{xG^4}^{\rm rad}\,.
\eea
In the latter equation (as well as in Eq. \eqref{deltap1brrG4}) the notation $P_{xG^4}^{\rm rad}\equiv [P_x]^{G^4}$ denotes the $G^4$ contribution to 
$P_x\equiv P_{\rm rad}{}_\mu e_x^\mu$ (without factoring masses or symmetric mass ratio), so that 
\beq
P_{xG^4}^{\rm rad}\equiv [P_x]^{G^4}=\frac{G^4m_1^2 m_2^2}{b^4}(m_2-m_1)\frac{h^4}{(\gamma^2-1)^2}\check P_{x}^4\,. 
\eeq

Using the explicit expression for $\Delta p_{1bG^4}^{\rm rr, lin-odd}$ given in Eqs. (12.37) and (12.38) of  Ref. \cite{Bini:2022enm}
we get from the first equation above the following TF-predictions for $c^{4 \rm diss}_{\rm 1b,1rad}$:
\bea
c^{4 \rm diss}_{\rm 1b,1rad}&=& \frac{G^4}{b^4}m_1^2 m_2^2 
\left\{(m_1+m_2)\left[
{\mathcal E}(\gamma)\frac{\g (6 \g^2 - 5)}{(\g^2 - 1)^{3/2}} \right.\right.\nonumber\\ 
&-&\left.  \pi \frac34 \hat J_2(\gamma) \frac{(5 \g^2 - 1)}{(\g^2 - 1)^{3/2}} -\hat J_3(\gamma) \frac{(2 \g^2 - 1)}{(\g^2 - 1)^2}\right]\nonumber\\
&-& \left. m_1 {\mathcal E}(\gamma) \frac{2\gamma^2-1}{(\gamma+1)\sqrt{\gamma^2-1}}\right\}\,.
\eea
In this relation we see that the TF formalism points out the presence of angular momentum losses of $G^2$ and $G^3$ orders in $c^{4 \rm diss}_{\rm 1b,1rad}$.
In particular, focusing on the 1GSF projection, i.e., the coefficient of $m_1^2 m_2^3$ of this equation, we can obtain an expression of $\hat J_3$ in terms of  
$c^{4 \rm diss}_{\rm 1b,1rad}{}^{\rm 1SF}$ together with the $G^2$ angular momentum loss and the $G^3$ energy loss (${\mathcal E}=\frac{\hat E_3}{(\gamma^2-1)^{3/2}}$). This yields 
\bea
\label{eqJ31}
\hat J_3 &=& C^{\rm ac} {\rm arccosh}(\gamma)+ C^{\rm at} {\rm at}\left(\frac{\sqrt{\gamma^2-1}}{\gamma}\right)\nonumber\\
&+& C^{\rm ln} \ln \left(\frac{\gamma+1}{2} \right)\,,
\eea
with
\bea
\label{eqJ32}
C^{\rm ac}&=& -\pi\frac{  \g \left(2 \g^2-3\right)}{8 \left(\g^2-1\right)^{3/2}} (35 \g^5-240 \g^4\nonumber\\
&&-70 \g^3+288 \g^2+19 \g-48 )\,, \nonumber\\
C^{\rm at}&=& -\pi\frac{3   \g \left(10 \g^6-27 \g^4+20 \g^2-3\right)}{\left(\g^2-1\right)^{3/2}}\,,\nonumber\\
C^{\rm ln}&=& \frac{1}{4} \pi   (35 \g^5-90 \g^4-70 \g^3\nonumber\\
&+& 16 \g^2+155 \g-62 )\,,
\eea
which indeed coincides with the result obtained in Ref. \cite{Manohar:2022dea}.

Finally, the second line in Eq. \eqref{c1and2rad} points out the existence of a simple proportionality relation between $c^{4 \rm diss}_{\rm 1b,2rad}$ and 
$P_{xG^4}^{\rm rad}$. This TF-predicted relation is indeed satisfied by the WEFT result for $c^{4 \rm diss}_{\rm 1b,2rad}$ of Ref. \cite{Dlapa:2022lmu}.

\section{TF-predicted structure of the $O(G^5)$  impulse}
\label{G5sect} 

At order $G^5$ the TF formalism predicts  that the impulse   contains again  three main contributions: a conservative one, a term linear in  $J_{\rm rad}$ and a term linear in $P_{\rm rad}$. However, at order $G^5$ we   have to take into account the presence of a {\it supplementary} contribution $\Delta  p_{1}^{ \rm rr \, remain \, sup}{}^\mu$, 
namely
\bea
\label{decoatG5}
\Delta p_a^\mu  &=&\Delta p_a^{{\rm cons}\,\mu}+\Delta  p_{a }^{{\rm rr  lin} J_{\rm rad}\, \mu}\nonumber\\
&+&\Delta  p_{a }^{{\rm rr  lin} P_{\rm rad}\mu }+\Delta  p_{a}^{ \rm rr \, remain \, sup}{}^\mu\,.
\eea
Separating out the supplementary contribution 
\beq
\label{eqremainsup2}
\Delta  p_{1}^{ \rm rr \, remain \, sup}{}^\mu=\frac{G^5}{b^5}m_1^2 m_2^3 f_b^{G^5 \rm remain }\hat b^\mu +O(G^6)\,,
\eeq
the linear-in-$P_{\rm rad}$ contribution at order $G^5$ is given by the rhs of Eq. \eqref{deltaplinP} where the  last term, $\Delta  p_{a}^{ {\rm rr \, remain}P_{\rm rad} }{}^\mu$,
was given in Ref. \cite{Bini:2022enm} as the sum of four contributions given in Eqs. (12.53), (12.62), (12.67) and (12.78) of Ref. \cite{Bini:2022enm}, describing its components along the basis $\hat b^\mu$, $(u_1^\mu+u_2^\mu)$ and $(u_1^\mu-u_2^\mu)$. Explicitly (using Eq.   \eqref{en12} below), 
\bea
\label{deltap1G5remain1}
\Delta  p_{1}^{G^5 {\rm rr \, remain}\, P_{\rm rad}\, \mu}&=&c_{1b}^{G^5 {\rm rr \, remain}  P_{\rm rad} }\hat b^\mu \nonumber\\
&+& c_{1+2}^{G^5 \rm rr \, remain }(u_1^\mu+u_2^\mu)\nonumber\\
&+& c_{1-2}^{G^5 \rm rr \, remain }(u_1^\mu-u_2^\mu)\,,
\eea
where 
\bea
\label{deltap1G5remain2}
c_{1b}^{G^5 {\rm rr \, remain} P_{\rm rad} } 
&=&  \frac{G^5}{b^5} m_1^3 
      m_2^3 (m_1 + m_2)^2  \frac{(\gamma+1)}{E^2} p_{xG^5}(\gamma) 
\,,\nonumber\\
c_{1+2}^{G^5 \rm rr \, remain }&=& 
\frac{G^5}{b^5} m_1^3 m_2^3 (m_1 - m_2) (m_1 + m_2)\times \nonumber\\
&&\frac{(2\gamma^2 - 1)}{E^2\sqrt{\gamma^2-1}} 
p_{xG^4}(\gamma)\,,
\nonumber\\
c_{1-2}^{G^5 \rm rr \, remain }&=& 
-2\frac{G^5}{b^5} m_1^4 m_2^4 \frac{(2\gamma^2-1)(\gamma+1)}{E^2\sqrt{\gamma^2-1}} 
p_{xG^4}(\gamma)\nonumber\\ 
&+& 
\frac{G^5}{b^5} m_1^3 m_2^3\frac{(2\gamma^2 - 1) (\gamma + 1)}{(\gamma - 1)\sqrt{\gamma^2-1}} 
p_{xG^4}(\gamma)\,.\qquad
\eea
Here we introduced the  notation   
\bea
[P_x]^{G^4}&\equiv & \frac{G^4m_1^2 m_2^2}{b^4} (m_2-m_1)p_{xG^4}(\gamma)\,,\nonumber\\
{}[P_x]^{G^5}&\equiv & \frac{G^5m_1^2 m_2^2(m_1+m_2) }{b^5}(m_2-m_1)p_{xG^5}(\gamma)\,.\qquad
\eea

Let us note that, when gathering both the $G^4$ and $G^5$ contributions, the remain-$P_{\rm rad}$ contribution to the impulse lies in the direction  of the $e_x^\mu$ vector and can be written in the simple form
\bea
\Delta p_{1x}^{{\rm rr \, remain} P_{\rm rad} } &=&  m_1^3 m_2^3 (m_1+m_2)  \frac{\gamma+1}{E^2}\left(\frac{G^4}{b^4} p_{xG^4}(\gamma)\right.\nonumber\\ 
&+&\left.\frac{G^5}{b^5}(m_1+m_2)p_{xG^5}(\gamma)\right)\nonumber\\
&=& m_1  m_2  (m_1+m_2)  \frac{\gamma+1}{E^2} \frac{P_{x}^{G^4}+P_{x}^{G^5}}{m_2-m_1}\,.\nonumber\\
\eea
We have explicitly checked that the TF predicted impulse, which in its original definition given by Eq. \eqref{finale},
for $\Delta p_1$ at 5PM  contains $1/E^2=1/(m_1^2+m_2^2+2\gamma m_1m_2)$ denominators (as well as $1/(m_1+m_2)^{2n}$ coming from the powers of $\nu$ in $\tilde \chi^{\rm cons}$, Eq. \eqref{eq_chi_tilde})  simplifies (because of the additional $\Delta  p_{1}^{G^5 \rm rr \, remain }$  terms) to a purely polynomial-in-masses expression of the type
\begin{widetext} 
\bea
\label{plefka_eqs}
\Delta p_1^{(5)\mu} &=& m_1 m_2 (m_2^4 \Delta p_{\rm 0SF}^{(5)\mu}[\gamma, \hat b^\mu  ,\check u_1^\mu, \check u_2^\mu ] + 
   m_1 m_2^3 \Delta p_{\rm 1SF}^{(5)\mu}[\gamma, \hat b^\mu , \check u_1^\mu, \check u_2^\mu] + 
   m_1^2 m_2^2 \Delta p_{\rm 2SF}^{(5)\mu} [\gamma,\hat b^\mu , \check u_1^\mu, \check u_2^\mu]\nonumber\\ 
&+&     
   m_1^3 m_2 \Delta p_{\overline{\rm 1SF}}^{(5)\mu}[\gamma, \hat b^\mu , \check u_1^\mu , \check u_2^\mu]
+m_1^4 \Delta p_{\overline{\rm 0SF}}^{(5)\mu} [\gamma, \hat b^\mu , \check u_1^\mu, \check u_2^\mu])\,,\nonumber\\
\Delta p_2^{(5)\mu} &=& m_1 m_2 (m_1^4 \Delta p_{\rm 0SF}^{(5)\mu}[\gamma, -\hat b^\mu  ,\check u_2^\mu, \check u_1^\mu ] + 
   m_2 m_1^3 \Delta p_{\rm 1SF}^{(5)\mu}[\gamma, -\hat b^\mu  ,\check u_2^\mu, \check u_1^\mu ]+ 
   m_1^2 m_2^2 \Delta p_{\rm 2SF}^{(5)\mu} [\gamma, -\hat b^\mu  ,\check u_2^\mu, \check u_1^\mu ]\nonumber\\ 
&+&    
   m_2^3 m_1 \Delta p_{\overline{\rm 1SF}}^{(5)\mu}[\gamma, -\hat b^\mu  ,\check u_2^\mu, \check u_1^\mu ]
+ m_2^4 \Delta p_{\overline{\rm 0SF}}^{(5)\mu}[\gamma, -\hat b^\mu  ,\check u_2^\mu, \check u_1^\mu ])\,.
\eea
\end{widetext}
Each $\Delta p_{\rm \check kSF}^{(5)\mu}[\gamma, \hat b^\mu  ,\check u_1^\mu, \check u_2^\mu ]$ is then decomposed along $\hat b^\mu$, $\check u_1^\mu$, $\check u_2^\mu$ according to the following notation
\bea
\label{plefka_eqs2}
\Delta p_{\rm \check kSF}^{(5)\mu}[\gamma, \hat b^\mu  ,\check u_1^\mu, \check u_2^\mu ]&=&
 \hat b^\mu  f_b^{G^5,\rm nSF}(\gamma) + \check u_1^\mu  f_{\check u_1}^{G^5,\rm nSF}(\gamma)\nonumber\\ 
&+& \check u_2^\mu  f_{\check u_2}^{G^5,\rm nSF}(\gamma)\,,
\eea
where $\check k=0,1,2,\bar 1, \bar 0$.
The 0SF and ${\overline {\rm 0SF}}$ contributions are defined by the probe limit and must satisfy the constraints 
\bea
f_b^{G^5,{\rm 0SF}}&=&f_b^{G^5,{\overline {\rm 0SF}}}\,,\qquad f_{\check u_1}^{\rm 0SF}=-f_{\check u_2}^{\overline{\rm 0SF}}\,,\nonumber\\
f_{G^5,\check u_2}^{\rm 0SF}&=& 0=f_{G^5,\check u_1}^{\overline{\rm 0SF}} \,. 
\eea
They are explicitly given in Table \ref{allfG5} of Appendix \ref{appendixex}.
 
In other words, the TF approach  predicts explicit forms for all the $\gamma$-dependent coefficients,
say $f_b^{\rm 0SF},\ldots f_b^{\rm 2SF}\,,\ldots $ entering the 5PM impulse, Eqs. \eqref{plefka_eqs}, in terms of the basic TF ingredients $\chi^{\rm cons}$, $P_\mu^{\rm rad}$ and the center-of-mass angular momentum loss $J_{\rm rad}$. 
The new ingredients entering at the $O(G^5)$ level are the following  nine functions of $\gamma$: $\tilde \chi_5^{1\rm cons}(\gamma)$, $\tilde \chi_5^{2\rm cons}(\gamma)$, $\hat J_4^0(\gamma)$, $\hat J_4^1(\gamma)$, $\hat P_{1+2}^{5,0}(\gamma)$,  $\hat P_{1+2}^{5,1}(\gamma)$, $\hat P_b^{5}(\gamma)$, $\hat P_{1-2}^{5}(\gamma)$ and $f_b^{G^5\, {\rm remain}}(\gamma)$.  
As discussed later, the latter quantity (which was defined in Eq. \eqref{eqremainsup})  is degenerate with an additional term in the $O(G^5)$ conservative scattering angle, $\tilde \chi_5^{\rm cons}$. The four, $P_{\rm rad}^\mu$-related, functions of $\gamma$, namely $\hat P_b^5(\gamma)$, $\hat P_{1-2}^5(\gamma)$, $P_{1+2}^{5,0}(\gamma)$ and $P_{1+2}^{5,1}(\gamma)$ are defined such that (see, e.g., Eqs. (11.8) in Ref. \cite{Bini:2022enm})
\bea
\label{Pb1p21m2G^5}
P_b^{G^5}  
&=& \frac{G^5 m_1^2 m_2^2 }{b^5}(m_2-m_1)(m_1+m_2)\hat P_b^5(\gamma)\,,\nonumber\\
P_{1-2}^{G^5}
&=& \frac{G^5 m_1^2 m_2^2 }{b^5}(m_2-m_1)(m_1+m_2)\hat P_{1-2}^5(\gamma)\,,\nonumber\\ 
P_{1+2}^{G^5}
&=& \frac{G^5 m_1^2 m_2^2 }{b^5}(m_1+m_2)^2  [\hat P_{1+2}^{5,0}(\gamma)+\nu \hat P_{1+2}^{5,1}(\gamma)]\,.\nonumber\\
\eea

Denoting generically with $K$ \lq\lq known terms," we find  at 1SF, ${\overline {\rm 1SF}}$ and 2SF levels the following results:
\bea
\label{f5TF1}
f_b^{G^5,{\rm 1SF}}&=& -\frac{2 \tilde \chi_5^{1\rm cons}}{(\gamma^2-1)^2}-\frac{\chi_1 \hat J_4^0}{b^5(\gamma^2-1)^2}+K \,,\nonumber\\
f_{\check u_1}^{G^5,{\rm 1SF}}&=& K \,, \nonumber\\
f_{\check u_2}^{G^5,{\rm 1SF}}&=& (\gamma-1)\hat P_{1-2}^5 +(\gamma+1)\hat P_{1+2}^{5,0}+K\,,
\eea
\bea
\label{f5TF2}
f_b^{G^5,{\overline {\rm 1SF}}}&= &   \hat P_b^5-\frac{2\tilde \chi_5^{\rm 1 cons}}{(\gamma^2-1)^2}-\frac{\hat J_{4}^0 \chi_1}{(\gamma^2-1)^2}  +K\,, \nonumber\\
f_{\check u_1}^{G^5,{\overline {\rm 1SF}}}&= & K\,, \nonumber\\
f_{\check u_2}^{G^5,{\overline {\rm 1SF}}}&= & -(\gamma-1)\hat P_{1-2}^5+(\gamma+1)\hat P_{1+2}^{5,0}+K \,,
\eea
and
\bea
\label{f5TF3}
f_b^{G^5,{\rm 2SF}}&=& f_b^{G^5\, {\rm remain}}+P_b^5-\frac{4}{(\gamma^2-1)^2}\tilde \chi_5^{1\rm cons} \nonumber\\
&+& \frac{2}{(\gamma^2-1)^2}\tilde \chi_5^{2\rm cons}\nonumber\\
&-& \frac{2\chi_1}{(\gamma^2-1)^2}\hat J_4^0 -\frac{\chi_1}{(\gamma^2-1)^2}\hat J_4^1   +K\,,\nonumber\\
f_{\check u_1}^{G^5,{\rm 2SF}}&= & K\,, \nonumber\\
f_{\check u_2}^{G^5,{\rm 2SF}}&= & (\gamma+1) \left(2\hat P_{1+2}^{5,0}+\hat P_{1+2}^{5,1}\right)+K\,.
\eea

These TF-predicted results satisfy the constraints coming from the linear momentum balance, namely
\beq
\label{balanceeq}
P_{\rm rad}^\mu=-\Delta p_1^\mu-\Delta p_2^\mu\,.
\eeq

This linear momentum balance implies the following constraints between the expansion coefficients of the impulse and those of the linear momentum loss
\bea
\label{Pb1p21m2G^5}
\hat P_b^5(\gamma)&=& f_b^{G^5, \rm 1SF} - f_b^{G^5,\overline {\rm 1SF}} \,,\nonumber\\
\hat P_{1-2}^5(\gamma)&=& \frac{1}{2(\gamma-1)}[f_{\check u_1}^{G^5,\rm 1SF} -f_{\check u_1}^{G^5,\overline {\rm 1SF}}\nonumber\\
& -&(f_{\check u_2}^{G^5,\rm 1SF} - 
f_{\check u_2}^{G^5,\overline {\rm 1SF}} )]\,,\nonumber\\
{}\hat P_{1+2}^{5,0}(\gamma)&=& \frac{1}{2(\gamma+1)}[f_{\check u_1}^{G^5,\rm 1SF}
+f_{\check u_1}^{G^5,\overline {\rm 1SF}}\nonumber\\
&+& f_{\check u_2}^{G^5,\rm 1SF} 
+f_{\check u_2}^{G^5,\overline {\rm 1SF}}]\,,\nonumber\\
{}\hat P_{1+2}^{5,1}(\gamma)&=& \frac{1}{\gamma+1}\left[\left(f_{\check u_1}^{G^5,\rm 2SF} 
+ f_{\check u_2}^{G^5,\rm 2SF}\right)\right.\nonumber\\ 
&-&\left. (f_{\check u_1}^{G^5,\rm 1SF}+f_{\check u_1}^{G^5,\overline {\rm 1SF}}
+f_{\check u_2}^{G^5,\rm 1SF} +f_{\check u_2}^{G^5,\overline {\rm 1SF}})
 \right]
\,.\nonumber\\
\eea
 
Remembering that at any given PM order the mass-shell constraint 
\beq
\label{masshelfine}
0=(p_1^{\rm out})^2-(p_1^{\rm in})^2=2p_1\cdot \Delta p_1 + (\Delta p_1)^2\,,
\eeq
determines the $\check u_1$ component of $\Delta p_1$ in terms of lower $G$ contributions (i.e., $f^{\rm nSF}_{\check u_1}=K$ at any PM order) we see that the relations \eqref{Pb1p21m2G^5} uniquely determine $f_{\check u_2}^{\rm 1SF}$ and $f_{\check u_2}^{\overline {\rm 1SF}}$ in terms of $P_{\rm rad}$, namely at $G^5$,
\bea
f_{\check u_2}^{G^5 \rm 1SF} &\sim & \hat P_{1-2}^{5,0} +\hat P_{1+2}^{5,0}+K\,, \nonumber\\
f_{\check u_2}^{G^5 \overline{\rm 1SF}} &\sim & \hat P_{1-2}^{5} +\hat P_{1+2}^{5,0} +K\,,\nonumber\\
f_{\check u_2}^{G^5 \rm 2SF} &\sim & \hat P_{1-2}^{5,0} +\hat P_{1+2}^{5,0}+\hat P_{1+2}^{5,1}+K\,, 
\eea
while they only determine the difference between $f_{b}^{G^5 \rm 1SF}$ and $f_{b}^{G^5 \overline{\rm 1SF}}$:
\beq
f_{b}^{G^5 \rm 1SF}-f_{b}^{G^5 \overline{\rm 1SF}} \sim   \hat P_b+K\,.
\eeq
On the other hand the balance equation \eqref{balanceeq} gives no constraints on $f_b^{G^5 \rm 2SF}$.

It is easily seen that the TF derived results \eqref{f5TF1}, \eqref{f5TF2} and  \eqref{f5TF3} are compatible with all those constraints. Note for instance how the same combination
\beq
-\frac{2 \tilde \chi_5^{1\rm cons}}{(\gamma^2-1)^2}-\frac{\chi_1 \hat J_4^0}{(\gamma^2-1)^2}
\eeq
enters both $f_b^{G^5 \rm 1SF} $ and $f_b^{G^5 \overline{\rm 1SF}}$.

\section{Tail-of-tail effects and the appearance of $p_\infty$-odd contributions in the conservative action at $G^5$}

As mentioned at the end of the introduction and summarized in Table \ref{table_loss},  
there is a progressive loss of $p_\infty$-parity property of the impulse as the PM order increases. In the previous section we showed how at order $G^4$ it was still possible to identify the conservative (and $p_\infty$-even) part of the dynamics in various formalisms, TF and WEFT, and to uniquely separate the radiation reaction part in  $p_\infty$-odd and  $p_\infty$-even parts.

\begin{figure*}
\includegraphics[scale=1.75]{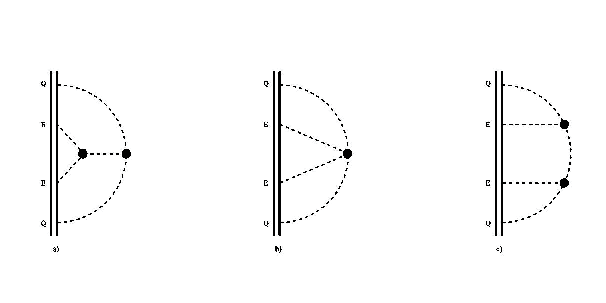}
\caption{\label{diagrams} Sketch of the various quadratic-tail and tail-of-tail  diagrammatic contributions to the nonlocal (radiation-zone-mediated)  effective action of a binary system. They start at order $G^5$ and get contributions from all multipole couplings (here indicated as $Q$).}
\end{figure*}

A new feature happens at order $G^5$ (see Fig. \ref{diagrams}) because of the arising of nonlinear hereditary (tail) effects. We recall that the presence of tail-transported hereditary (nonlocal in time) effects in binary dynamics was first pointed out in Ref. \cite{Blanchet:1987wq}. Such effects are included in the MPM formalism when solving 
for the metric at higher PM orders \cite{Blanchet:1992br}. Tail effects, related to the scattering of gravitational waves by the total energy $E$ of the binary system,  arise at various orders in $GE$. Linear tail effects include only one power of $GE$ (and first arise in the binary dynamics at the $G^4/c^8$ level, i.e., 4PM and 4PN, see \cite{Blanchet:1987wq,Foffa:2011np,Damour:2014jta}). Quadratically  nonlinear tail effects \cite{Blanchet:1992br,Blanchet:1997jj} (henceforth called tail-of-tail, tail-squared  or quadratic-tail) involve two powers of $(GE)$.
They first arise in the binary dynamics at the 
$G^5/c^{11}$ level, i.e., 5PM and 5.5 PN as was progressively understood in Refs.  \cite{Shah:2013uya,Bini:2013rfa,Blanchet:2013txa,Blanchet:2014bza}. The conservative part of the binary dynamics mediated by quadratic-tail hereditary  effects was described at 5.5PN order by an action  involving \lq\lq split" bilinear expressions in the (even parity) quadrupole moment of the binary system   \cite{Damour:2015isa,Bini:2020wpo}, see Eq. (9.19) in Ref. \cite{Damour:2015isa} and Eq.  (12.2) in Ref. \cite{Bini:2020wpo} 
and Eq.  (12.2) in Ref. \cite{Bini:2020wpo}. See also 
Refs. \cite{Edison:2022cdu,Edison:2023qvg} for a derivation of the corresponding
quadrupolar tail-of-tail, Schwinger-Keldysh (doubled-variable) non-conservative action.

Generalizing the derivation of the quadratic-tail, even-quadrupole conservative action of Refs. \cite{Damour:2015isa,Bini:2020wpo} to general-$l$ (even-parity and odd-parity) multipoles,  we get the following {\it conservative} quadratic-tail, or \lq\lq tail-of-tail,"  nonlocal action. 
This action can be written (modulo integration by parts) in two forms: a manifestly time-symmetric form involving the logarithmic kernel $\ln \frac{c|t-t'|}{2r_0}$ (where $r_0$ is an arbitrary scale which actually does not contribute to the nonlocal part of the action),
namely
\bea
\label{eqtailoftail0}
S_{\rm time-sym}^{\rm tail-of-tail}&=&\frac12 \left(\frac{G{\mathcal M}}{c^3}\right)^2G \sum_{l\ge 2} a_l \beta_l^{\rm even}\int dt \times \nonumber\\
&&\int_{-\infty}^\infty dt' I_L^{(l+2)}(t)I_L^{(l+2)}(t')\ln \frac{c|t-t'|}{2r_0}\nonumber\\
&+& \frac12 \left(\frac{G{\mathcal M}}{c^3}\right)^2G \sum_{l\ge 2} b_l \beta_l^{\rm odd}\int dt \times \nonumber\\
&&\int_{-\infty}^\infty dt' J_L^{(l+2)}(t)J_L^{(l+2)}(t')\ln \frac{c|t-t'|}{2r_0}\,,\nonumber\\
\eea
or a non-manifestly time-symmetric (\lq\lq time-split") form involving a scale-free principal-value integral, namely
\bea
\label{eqtailoftail}
S^{{\rm tail-of-tail}}_{\rm time-split}&=& \frac14  \left(\frac{G{\mathcal M}}{c^3}\right)^2G \int dt \int _{-\infty}^\infty \frac{d \tau}{\tau} {\mathcal G}^{\rm split}(t,\tau)\,,\qquad
\eea
with
\beq
\label{actionbetagen}
{\mathcal G}^{\rm split}(t,\tau)=\sum_{l\ge 2}\left( a_l \beta_l^{\rm (even)}{\mathcal G}_I^l(t,\tau)
+b_l \beta_l^{\rm (odd)}{\mathcal G}_J^l(t,\tau)\right)\,,
\eeq 
where
\bea
{\mathcal G}_I^l(t,\tau)&\equiv &I_L^{(l+1)}(t)[I_L^{(l+2)}(t+\tau)-I_L^{(l+2)}(t-\tau)]\,, \nonumber\\
{\mathcal G}_J^l(t,\tau)&\equiv & J_L^{(l+1)}(t)[J_L^{(l+2)}(t+\tau)-J_L^{(l+2)}(t-\tau)]\,.\qquad
\eea
Here, ${\mathcal M}\equiv E/c^2$, 
 $I_L$ and $J_L$ are the even- and odd-parity source multipole moments (as defined in the Multipolar Post-Minkowskian formalism), and $a_l$ and $b_l$ denote the coefficients entering the even and odd multipolar parts of the energy flux~\cite{Blanchet:2013haa}
\bea
a_l&=&\frac{(l+1)(l+2)}{l(l-1)l! (2l+1)!!}\,,\nonumber\\ 
b_l&=&\frac{4 l (l+2)}{(l-1)(l+1)! (2l+1)!!}\,.
\eea

In addition  $\beta_l^{\rm (even,odd)}$ are the beta function coefficients describing the scale dependence of the multipole moments. $\beta_l^{\rm (even)}$ was first computed in the Appendix of Ref. \cite{Blanchet:1987wq}; its meaning  as a coefficient  related to scale-dependence was then pointed out in   Ref. \cite{Blanchet:1997jj} and understood as a renormalization-group $\beta$ coefficient in Ref. \cite{Goldberger:2009qd}. According to recent works \cite{Fucito:2024wlg,Ivanov:2025ozg}  the odd-parity $\beta_l^{\rm (odd)}$'s are equal to their even-parity counterparts. 

We recall that linear-tail conservative (time-symmetric) nonlocal actions are logarithmically UV-divergent and 
must be completed by corresponding nonlocal, time-antisymmetric  radiation-reaction terms which are UV-finite (see
Sec. VI of \cite{Damour:2014jta}). By contrast, the
 quadratic-tail conservative nonlocal actions, Eqs. \eqref{eqtailoftail0} and \eqref{eqtailoftail}, are UV-finite at $t'-t=\tau=0$, and must be completed by  corresponding  (in-out) multipolar radiation-reaction terms which are logarithmically divergent \footnote{The latter, external-zone, logarithmic UV-divergences will be compensated by corresponding IR-divergences in the matched source multipole moments.}. 

The leading PN order value of the action \eqref{actionbetagen}  in Refs. \cite{Damour:2015isa,Bini:2020wpo}
is obtained by using $\beta_2=-\frac{214}{105}$ and $a_2=\frac15$ and reads, with $B=-\frac{107}{105}=\frac{\beta_2}{2}$,
\bea
\label{ouraction}
S^{{\rm tail-of-tail}}_{\rm quadrupolar}&=& -\frac{B}{10} \frac{G}{c^5}\left(\frac{G {\mathcal M}}{c^3} \right)^2 \int  dt \times \nonumber\\
&&I_{ij}^{(3)}(t)\int _{-\infty}^\infty \frac{d\tau}{\tau}[I_{ij}^{(4)}(t+\tau)-I_{ij}^{(4)}(t-\tau))]\nonumber\\
&=&  -\frac{B}{10} \frac{G}{c^5}\left(\frac{G {\mathcal M}}{c^3} \right)^2 \int dt \int _{-\infty}^\infty \frac{d\tau}{\tau} {\mathcal G}_I^2(t,\tau)
\,.\nonumber\\
\eea
 
The 5.5PN and $O(G^5)$ level contribution to the conservative scattering angle induced by  Eq. \eqref{ouraction} was computed in Ref. \cite{Bini:2020hmy}, see Eq. (5.10) there,
and found to be 
\beq
\label{chi5p5}
\chi_{{\rm 5.5}}=- \nu \frac{95872}{675}\frac{p_\infty^6}{j^5}+O(G^6)\,.
\eeq
We will prove below that this term, together with several other TF-computed terms, is included in the recent $G^5$ 1SF result of Ref. \cite{Driesse:2024feo}.

\section{Comparison between the $O(G^5)$ TF impulse and the 1SF WEFT results}

We have presented in Section IV above the TF ingredients entering the 1SF $O(G^5)$ impulse. Namely, from  Eqs. \eqref{f5TF1} and \eqref{f5TF2}, 

\begin{enumerate}
  \item The $O(\nu^1)$ contribution to the 5PM conservative scattering angle: $\tilde \chi_5^{1\rm cons}(\gamma)$;
  \item The $O(\nu^0)$ contribution to the 4PM angular momentum loss: $\hat J_4^0(\gamma)$;
  \item The $O(G^5)$ (mass-factorized) component of $P_{\rm rad}^\mu$ along $u_1^\mu-u_2^\mu$: $\hat P^5_{1-2}(\gamma)$;
  \item The $O(G^5)$ (mass-factorized)  component of $P_{\rm rad}^\mu$ along $\hat b^\mu$: $\hat P^5_{b}(\gamma)$;
  \item The $O(\nu^0)$, $O(G^5)$ (mass-factorized)  component of $P_{\rm rad}^\mu$ along $u_1^\mu+u_2^\mu$: $\hat P^{5,0}_{1+2}(\gamma)$. 
\end{enumerate}

In other words, a PN-exact comparison between TF and WEFT at 1SF would require the knowledge of five independent functions of $\gamma$. However,
the TF formalism gives only access to the  PN expansion of those five functions.
The current PN knowledge of these functions in the TF formalism is the following: 
\begin{enumerate}
  \item   $\tilde \chi_5^{1\rm cons}(\gamma)$ has been computed in Refs. \cite{Bini:2020wpo,Bini:2020nsb,Bini:2020hmy,Bini:2020rzn} up to the absolute 6PN accuracy (including the tail-of-tail 5.5PN contribution highlighted in the previous section), 
\bea
\tilde \chi_{5\, {\rm TF}}^{1\rm cons} &=&  \frac{2}{5p_\infty^3}-\frac{121}{10 p_\infty}+\left(-\frac{19457}{60}+\frac{41}{8}\pi^2\right) p_\infty\nonumber\\
&+&\left(-\frac{5211479}{4320}+\frac{5069}{144}\pi^2-\frac{6272}{45}\ln(2p_\infty)\right) p_\infty^3\nonumber\\
&+&\left(-\frac{782142451}{504000}+\frac{111049}{960}\pi^2 \right.\nonumber\\
&-&\left.\frac{74432}{525}\ln(2 p_\infty)\right)p_\infty^5\nonumber\\
&-& \frac{1}{2} \frac{95872}{675} p_\infty^6 \nonumber\\
&+& \left(\frac{57623004613}{98784000}-\frac{184881}{4480}\pi^2\right.\nonumber\\
&-&\left. \frac{881392}{11025}\ln(2p_\infty)\right)p_\infty^7\nonumber\\
&+& O(p_\infty^8)\,,
\eea
where the extra factor $1/2$ in the 5.5PN tail-of-tail contribution  comes from the definition of $\tilde \chi_n$, Eq. \eqref{chi_cons_def}.  

\item  $\hat J_4^0(\gamma)$ was computed in Ref. \cite{Bini:2022enm}  up to the 5.5PN absolute accuracy (fractional 3PN), namely
\bea
\hat J_{4\rm TF}^0(\gamma)&=& \frac{176}{5}p_\infty+\frac{8144}{105} p_\infty^3+\frac{448}{5} p_\infty^4 \nonumber\\
&-&\frac{93664}{1575} p_\infty^5
+\frac{1184}{21}p_\infty^6
-\frac{4955072}{121275}p_\infty^7\nonumber\\
&+&O(p_\infty^8)\,.
\eea
In addition, Ref. \cite{Heissenberg:2025ocy} recently computed the even-in-$p_\infty$ part of $\hat J_4^0$ to all PN orders, with a result whose PN expansion reads
\bea
\hat J_{4\rm even}^0&=&\frac{448}{5} p_\infty^4  + \left(\frac{1184}{21}  - \frac{431936}{1575} \nu\right) p_\infty^6\nonumber\\ 
&+& \left(- \frac{13736}{315}  -  \frac{126736 \nu}{1575} \right.\nonumber\\ 
&+&\left.  \frac{857152}{1575} \nu^2  \right) p_\infty^8 \nonumber\\
&+& \left(\frac{724868}{17325}  +  \frac{3100472}{121275} \nu \right.\nonumber\\   
&+&\left.  \frac{4544}{315} \nu^2    -  \frac{472256}{525 } \nu^3  \right) p_\infty^{10}
+O(p_\infty^{12})\,.
\eea
Therefore, we can combine the two results to reach the (absolute) 6PN accuracy in the knowledge of $\hat J_{4\rm TF}^0(\gamma)$, namely
\bea
\label{J40combined}
\hat J_{4\rm combined}^0(\gamma) &=& \frac{176}{5}p_\infty+\frac{8144}{105} p_\infty^3+\frac{448}{5} p_\infty^4\nonumber\\
&-&\frac{93664}{1575} p_\infty^5 +\frac{1184}{21}p_\infty^6\nonumber\\
&-& \frac{4955072}{121275}p_\infty^7
- \frac{13736}{315} p_\infty^8\nonumber\\ 
&+& O(p_\infty^9)\,.
\eea
  
\item  $\hat P^5_{1-2}(\gamma)$ was computed in Ref. \cite{Bini:2022enm}  up to the 5.5PN absolute accuracy (fractional 3PN), 
namely
\bea
\label{P1m2G5}
\hat P_{1-2}^{5}(\gamma)&=&  \frac{82 \pi }{15  p_\infty^3}- \frac{5207 \pi }{630  p_\infty}+\frac{939 \pi ^3}{280}\nonumber\\ 
&-&\frac{1491 \pi }{400 }-\frac{963239 \pi}{40320 }  p_\infty\nonumber\\
&-& \left(\frac{13603 \pi ^3}{4480}-\frac{902743 \pi }{33600 }\right) p_\infty^2\nonumber\\
&-&  \left(\frac{1591 \pi  }{980 }\log
   (\frac{p_\infty}{2})-\frac{313 \pi ^3}{140 }\right.\nonumber\\
&+&\left. \frac{4809573323 \pi }{434649600 }\right)p_\infty^3\nonumber\\
&+& O(p_\infty^4)\,.
\eea

\item  $\hat P^5_{b}(\gamma)$ was computed in \cite{Bini:2022enm} up to the  5.5PN absolute accuracy (fractional 3PN), namely 
\bea
\label{P5bgamma}
\hat P^5_{b}(\gamma)&=& -\frac{64}{3 p_\infty^2}-\frac{37 \pi ^2}{20}-\frac{27392}{525}-\frac{30208}{225} p_\infty\nonumber\\
&+&\left(-\frac{856768}{33075}-\frac{3429\pi ^2}{1120}\right) p_\infty^2
+\frac{462592 }{7875}p_\infty^3\nonumber\\
&+&\left(-\frac{74417152}{363825}-\frac{7915 \pi ^2}{2688}\right) p_\infty^4\nonumber\\
&+& O(p_\infty^5)\,.
\eea

\item  $\hat P^{5,0}_{1+2}(\gamma)$ was computed in \cite{Bini:2022enm}  up to the 5.5PN absolute accuracy (fractional 3PN), namely 
\bea
\label{P1p250}
\hat P_{1+2}^{5,0}(\gamma)&=&   \frac{61 \pi }{5 p_\infty^3}+ \frac{34073 \pi }{1680  }\frac{1}{p_\infty}+\frac{297 \pi ^3}{40 }
 -\frac{23923 \pi }{2880}  p_\infty \nonumber\\
&+& \left(-\frac{31029 \pi ^3}{2240  }+\frac{1484997 \pi }{11200  }\right)p_\infty^2 \nonumber\\
&+&\left(-\frac{10593 \pi  }{700  }\log (\frac{p_\infty}{2})+\frac{99 \pi ^3}{20
}\right.\nonumber\\
&+& \left.\frac{34695068413 \pi }{620928000  }\right)p_\infty^3 \nonumber\\
&+&O(p_\infty^4)\,. 
\eea 

\end{enumerate}

Injecting the 6PN accurate results for $\tilde \chi_5^{\rm TF}$ and for $\hat J_{4\rm combined}^0(\gamma)$ in Eq. \eqref{f5TF1}
we found perfect agreement for $f_b^{G^5 \rm 1SF}$ with
the (1SF, WEFT) results  of Ref. \cite{Driesse:2024feo}.

This agreement is remarkable in view of the presence of both $T$-even and $T$-odd contributions  in $\chi_5^{1 \rm cons}$ and  $\hat J_{4\rm combined}^0(\gamma)$.
The $T$-odd contribution  in $\chi_5^{1 \rm cons}$ is linked to the tail-of-tail effects discussed above, while the $T$-odd terms in  $\hat J_4^0$ come from 
time-asymmetric contributions to the angular momentum loss. 

Ref. \cite{Driesse:2024feo} has computed the all  PN order values of $\hat P_{1-2}^5$, $\hat P_{b}^5$, $\hat P_{1+2}^{5,0}$. The latter three observables are respectively denoted in \cite{Driesse:2024feo} $r_{1}$, $r_2$ and $r_3$.
The TF-computed high-PN expansion values of $\hat P_{1-2}^5$, $\hat P_{b}^5$, $\hat P_{1+2}^{5,0}$ displayed above nicely agree, within their accuracy, with  the results of  \cite{Driesse:2024feo}. %Evidently, the latter results allow to complete the above PN expansion to any wished PN accuracy. 
Evidently, one can use the latter results to complete the TF-computed results for those three observables to any needed accuracy. For instance,   
Ref. \cite{Driesse:2024feo}  allows us to get the next 6PN level terms in  
\bea
\hat P^5_{b}|_{\rm 6PN}&=&-\frac{99136 }{2625}p_\infty^5\,,\nonumber\\
\hat P_{1-2}^{5}|_{\rm 6PN}&=&
\left( \frac{4118497 \pi}{44800} - \frac{40459\pi^3}{3584}  \right)p_\infty^4\,,\nonumber\\
\label{P1p250bis}
\hat P_{1+2}^{5, 0}|_{\rm 6PN}&=&
\left(\frac{100349 \pi ^3}{17920}-\frac{1815089 \pi }{26880}\right)p_\infty^4\,.
\eea

Let us finally emphasize that the TF formalism predicts the hidden presence of the angular momentum loss $\hat J_4^0$ and $\hat J_4^1$ in the $b$-component of the impulse, see Eqs. \eqref{f5TF1}, \eqref{f5TF2}, \eqref{f5TF3}. However, it comes together with the $G^5$ conservative scattering angle $\tilde\chi_5^{1 \rm cons}$. At order $G^4$ we could explicitly check the hidden presence of $J_{\rm rad}$ because $\tilde \chi_4^{\rm cons}$ was separately known, see section \ref{G4sect} above.
 
As soon as the WEFT formalism will compute the all-PN-order value of $\hat J_4^0(\gamma)$ then one will be able to read off the exact value  $\tilde \chi_5^{1\,{\rm cons}}$ from the current 1SF $G^5$ results of Ref. \cite{Driesse:2024feo}.

\section{Predictions for the 2SF-level impulse}

We have presented in Section IV above the TF ingredients entering the 2SF $O(G^5)$ impulse. Namely, from  Eqs. \eqref{f5TF3} 

\begin{enumerate}
  \item The $O(\nu^1)$ and $O(\nu^2)$ contributions to the 5PM conservative scattering angle: $\tilde \chi_5^{1\rm cons}(\gamma)$, $\tilde \chi_5^{2\rm cons}(\gamma)$;
  \item The $O(\nu^0)$ and $O(\nu^1)$ contributions to the 4PM angular momentum loss: $\hat J_4^0(\gamma)$ and $\hat J_4^1(\gamma)$;
  \item The $O(G^5)$ component of $P_{\rm rad}^\mu$ along $\hat b^\mu$: $\hat P^5_{b}(\gamma)$;
  \item The $O(\nu^0)$ and $O(\nu^1)$, $O(G^5)$ component of $P_{\rm rad}^\mu$ along $u_1^\mu+u_2^\mu$: $\hat P^{5,0}_{1+2}(\gamma)$ and $\hat P^{5,1}_{1+2}(\gamma)$;
  \item The supplementary contribution to $\Delta p_1^{\rm remain}$: $f_b^{G^5\, {\rm remain}}(\gamma)$.
\end{enumerate}

In other words, a PN-exact comparison between TF and WEFT at 2SF would require (beyond what was needed at 1SF) the knowledge of three independent functions of $\gamma$, namely $\hat J_4^1(\gamma)$, $\hat P^{5,1}_{1+2}(\gamma)$, $\tilde \chi_5^{2\, \rm cons}(\gamma)$.
However,
the TF formalism gives only access to   the PN expansion of those three functions.
Namely, the current PN knowledge of these functions in the TF formalism is the following:
\begin{enumerate}
  \item   $\tilde \chi_5^{2\rm cons}(\gamma)$ has been computed in Refs. \cite{Bini:2020wpo,Bini:2020nsb,Bini:2020hmy,Bini:2020rzn} up to the absolute 6PN accuracy and reads  
\bea
\tilde \chi_5^{2\,{\rm cons}}(\gamma)&=& \frac{1}{5}\frac{1}{p_\infty}+\frac{59}{10}  p_\infty
+\left(\frac{10681}{144}-\frac{41}{24}\pi^2\right) p_\infty^3\nonumber\\
&+& \left(\frac{1408}{45} \ln(2p_\infty) -\frac{4}{15}  \bar d_5^{\nu^2}-\frac{365555}{6048}\right.\nonumber\\
&+&\left.\frac{23407}{5760}\pi^2\right)p_\infty^5\nonumber\\
&+& \left(-\frac{4}{35} q_{4,5}^{\nu^2}-\frac{2273170241}{2352000}+\frac{1219303}{20160}\pi^2\right.\nonumber\\
&-&\left. \frac{6944}{225}\ln(2p_\infty)\right) p_\infty^7
+O(p_\infty^8)\,.
\eea
\item $\hat J_4^1(\gamma)$ was computed in Ref. \cite{Bini:2022enm} up to the 5.5PN absolute accuracy $O(p_\infty^7)$ (fractional 3PN), and its knowledge can be  raised to the 6PN ($O(p_\infty^8)$) level by using  the results of Ref. \cite{Heissenberg:2025ocy}. This yields the combined result
\bea
\hat J_4^1(\gamma) &=&  -\frac{208}{15}  p_\infty^3 
+ \frac{988}{63}  p_\infty^5
-\frac{13312}{525} p_\infty^6\nonumber\\
&+& \frac{5458}{1575}  p_\infty^7
-\frac{208}{225}  p_\infty^8+O(p_\infty^9)\,.
\eea
%%%%%%%%%%%%%%%%%%%%%55
\item $\hat P^{5,1}_{1+2}(\gamma)$ was computed in Ref. \cite{Bini:2022enm} up to the 5.5PN absolute accuracy (fractional 3PN), 
see below.

\end{enumerate}
Concerning the supplementary remain contribution, $f_b^{G^5\, {\rm remain}}(\gamma)$, it was pointed out  in Ref. \cite{Bini:2022enm} that it  could start at the 5PN level, i.e., its PN expansion would  a priori read
\bea
\label{fb_exp}
f_b^{G^5\,{\rm remain}}(\gamma)&=&f_1^{G^5, {\rm sup}} p_\infty +f_2^{G^5, {\rm sup}} p_\infty^2\nonumber\\
&+& f_3^{G^5, {\rm sup}} p_\infty^3+O(p_\infty^4)\,.
\eea
We see from   Eqs. \eqref{f5TF3}  that the supplementary $f_b^{G^5\, {\rm remain}}(\gamma)$ term is degenerate with $\tilde \chi_5^{2\rm cons}(\gamma)$ and can formally be absorbed in a redefinition of the $\tilde \chi_5^{2\rm cons}(\gamma)$ as follows
\beq
\tilde \chi_5^{2\rm cons}{}'(\gamma)=\tilde \chi_5^{2\rm cons}(\gamma)+\frac12 p_\infty^4 f_b^{G^5}(\gamma)\,.
\eeq
However, from the TF point of view, the $p_\infty$ structure of $\tilde \chi_5^{2\rm cons}(\gamma)$ is not arbitrary but is predicted by the structure of the local-in-time and nonlocal-in-time contributions to the dynamics. On the one hand, one expects the local-in-time dynamics to be $T$-even (and leading to a $p_\infty$-even conservative scattering angle). On the other hand, the nonlocal-in-time contribution to the conservative dynamics has a $p_\infty$-character dictated 
by the structure of the tail-mediated interactions. We saw above that tail-squared effects, connected with the action \eqref{ouraction}, lead to $p_\infty$-odd contributions to the scattering angle. However, at the 5.5PN level (corresponding to the term $f_2^{G^5, {\rm sup}} p_\infty^2$ in Eq. \eqref{fb_exp}) we have seen above that 
the  conservative scattering angle is linear in $\nu$ and therefore does not contribute to the presently considered 2SF level in $f_b^{G^5}$.

Finally, at the 2SF-level, the TF approach makes  predictions for the beginning of the PN expansion of two independent functions of $\gamma$: $f_b^{G^5, {\rm 2SF}}$ and $\hat P_{1+2}^{5,1}(\gamma)$ (the latter corresponding to the quantity denoted as $r_4$ in  \cite{Driesse:2024feo}). 
The value of $\hat P_{1+2}^{5,1}(\gamma)$  has been computed to the absolute 5.5PN level in Ref.  \cite{Bini:2022enm}  and reads  
\bea
\hat P_{1+2}^{5,1}(\gamma)&=&  -\frac{55 \pi }{12 } \frac{1}{p_\infty} 
+ \frac{6427 \pi  }{10080  }p_\infty\nonumber\\
&+&\left(-\frac{939 \pi ^3  }{560}+\frac{877 \pi  }{400  } \right)p_\infty^2 \nonumber\\
&+&\left(-\frac{4059 \pi ^3  }{1280 }+\frac{255491 \pi }{10080 } \right)p_\infty^3 \nonumber\\
&+&\left( \frac{\pi E_{5,9}^1}{2}-\frac{190387 \pi ^3  }{4480}+\frac{14186677 \pi  }{33600} \right)p_\infty^4\nonumber\\
&+& O(p_\infty^5)\,.\nonumber\\
\eea
The presently known 5.5 PN level knowledge of $\hat P_{1+2}^{5,1}(\gamma)$ corresponds to the first four terms in the latter equation, up to the term $\propto  p_\infty^3$.
For information, we parametrized the next, 6PN-level term,   $\propto  p_\infty^4$ in the latter equation, by the (currently unknown) coefficient $E_{5,9}^1$ entering the $O(\nu)$ term in $E_5=E_5(\gamma,\nu)$ defined from 
\bea
E_{\rm rad}&=&M\nu^2 \left(\frac{G^3 E_3}{j^3}+\frac{G^4 E_4}{j^4}+\frac{G^5 E_5}{j^5}+\ldots\right)\,,
\eea
with (denoting below the fractional PN accuracy such that nPNabs=nPNfrac+2.5PN)
\beq
\label{E5_param}
E_5=E_5^{\le {\rm 3PNfrac}}+\left(E_{5,9}^0 + E_{5,9}^1 \nu +E_{5,9}^2 \nu^2\right)p_\infty^9+O(p_\infty^{10})\,,\qquad
\eeq
where
\bea
&&E_5^{\le {\rm 3PNfrac}} =\pi\left\{ \frac{122}{5} p_\infty^2
+\left(\frac{13831}{280}-\frac{933\nu }{10}\right) p_\infty^4\right.\nonumber\\
&+&\frac{297 \pi ^2}{20} p_\infty^5 \nonumber\\
&+&\left(\frac{2067\nu ^2}{10}-\frac{187559\nu }{1680}-\frac{64579}{5040}\right) p_\infty^6\nonumber\\
&+&\left[\left(-\frac{15291 \pi^2}{280}+\frac{296}{25}\right)\nu-\frac{24993 \pi^2}{1120}+\frac{9216}{35}\right] p_\infty^7 \nonumber\\
&+&\left[-\frac{1823\nu ^3}{5}+\frac{12269\nu ^2}{80}+\left(-\frac{4059 \pi^2}{640}+\frac{76897}{480}\right)\nu \right.\nonumber\\
&-&\left. \left.\frac{10593}{350}  \log (\frac{p_\infty}{2})+\frac{99 \pi ^2}{10}+\frac{29573617463}{310464000}\right]p_\infty^8\right\}\,,\qquad
\eea
as can be read from Eqs. (C10)-(C13) of Ref. \cite{Bini:2021gat} up to the 2PNfrac level, and from Eqs. (6.5) and (6.6) of Ref. \cite{Bini:2022enm} for the contributions to 2.5PNfrac and 3PNfrac, respectively.
The first ($\nu^0$) and the last ($\nu^2$) contributions to the  next, 3.5PNfrac, $O(p_\infty^9)$ term  (as written in parametric form in Eq.  \eqref{E5_param})
are given by
\bea
E_{5,9}^0&=& \left(\frac{1577 \pi ^2}{1120}-\frac{384}{7}\right)\,,\nonumber\\
E_{5,9}^2&=& \left(\frac{33399 \pi ^2}{280}-\frac{888}{25}\right)\,.
\eea
Here $E_{5,9}^0$ is known from Ref. \cite{Driesse:2024feo}, while $E_{5,9}^2$ follows  by using the $\nu$-polynomiality property \eqref{hatEn} 
\beq
\label{eqE5new}
E_5=\frac{\hat E_5}{h^6}=\frac{\hat E_5^0+\nu \hat E_5^1}{(1+2\nu (\gamma-1))^3}\,,
\eeq 
which implies   that all the powers of $\nu$ starting at $\nu^2$ in $E_5$  
come from PN-expanding the denominator in Eq. \eqref{eqE5new}.

The value of $f_b^{G^5, {\rm 2SF}}$ is predicted up to the absolute 6PN level and reads (using the corresponding PN expressions for 
$\hat J_2$, $\hat J_3$, $\hat J_4^0$, $\hat J_4^1$, $\hat P^3_{1+2}$, $\hat P^4_{1+2}$,$\hat P^4_{1-2}$, $\hat P_b^5$, $\chi_1$, $\tilde \chi^{\rm cons}_2$, $\tilde \chi^{0\, \rm cons}_3$, $\tilde \chi^{1\, \rm cons}_3$, $\tilde \chi^{0\, \rm cons}_5$,$\tilde \chi^{1\, \rm cons}_5$, $\tilde \chi^{2\, \rm cons}_5$)
\bea
\label{eqfb2SF}
f_{b\,{\rm TF}}^{G^5, {\rm 2SF}}&=&f_b^{G^5, {\rm 2SF,4.5PNabd} }\nonumber\\
&+& C_{b1}(d_5^{\nu^2},f_1^{G^5, {\rm sup}})p_\infty\nonumber\\
&+& C_{b2}(f_2^{G^5, {\rm sup}})p_\infty^2\nonumber\\
&+& C_{b3}(q_{4,5}^{\nu^2},f_3^{G^5, {\rm sup}})p_\infty^3+O(p_\infty^4)\,.
\eea
The 4.5PNabs-accurate TF contribution to $f_b^{G^5, {\rm 2SF}}$ is fully predicted to have the following value
\bea
\label{fbfinal}
&&f_{b\, {\rm TF}}^{G^5, {\rm 2SF,4.5PNabs} }=
-\frac{12}{p_\infty^9}+\frac{64}{p_\infty^7}+\frac{496+54 \pi ^2}{p_\infty^5}\nonumber\\
&+&\frac{832}{45 p_\infty^4}+\left(\frac{1552}{3}+\frac{481 \pi
   ^2}{2}\right)\frac{1}{p_\infty^3}\nonumber\\
&+& \left(\frac{409888}{1575}-\frac{94 \pi ^2}{5}\right)\frac{1}{p_\infty^2}\nonumber\\
&+& \left(\frac{25088}{45} \log (2 p_\infty)+\frac{21131 \pi ^2}{72}+\frac{190576}{135}\right)\frac{1}{p_\infty}\nonumber\\
&-&\frac{1489 \pi ^2}{35}+\frac{2283544}{11025}\,.
\eea
Beyond the $O(p_\infty^0)$ level there enter several  dimensionless parameters that are left undetermined by the TF formalism.
On the one hand,  the parameters $f_1^{G^5, {\rm sup}}$, $f_2^{G^5, {\rm sup}}$,  $f_3^{G^5, {\rm sup}}$ parametrizing $
f_b^{G^5\, {\rm remain}}(\gamma)$, Eq. \eqref{fb_exp}, enter at the absolute
 5PN, 5.5PN and 6PN levels, respectively.
On the other hand, the parameters $\bar d_5^{\nu^2}$ and $q_{4,5}^{\nu^2}$, which  respectively belong to the absolute 5PN and 6PN levels, enter the above expression in the form
\beq
\sim (\bar d_5^{\nu^2}+...)p_\infty +( q_{4,5}^{\nu^2}+...)p_\infty^3\,.
\eeq
Though the full value of $\bar d_5^{\nu^2}$ has not been yet determined,  the work of Refs. \cite{Blumlein:2020pyo,Blumlein:2021txe} (see also Eqs. 9.42 in \cite{Bini:2021gat}) has determined its transcendental contribution (which comes from the exchange of potential gravitons), namely
\beq
\bar d_5^{\nu^2}=\frac{306545}{512}\pi^2+d_5^{\rm rat}\,,
\eeq
where $d_5^{\rm rat}$ is a rational number.
Inserting the latter result in Eq. \eqref{eqfb2SF} above leads to the following TF-based prediction for the $O(p_\infty^1)$ (5PNabs) coefficient $C_{b1}$ entering the $O(p_\infty^1)$ contribution to $f_b^{G^5\, {\rm 2SF}}(\gamma)$
\bea
\label{eqpinf1}
C_{b1}&=&\frac{8}{15}d_5^{\rm rat}+f_1^{G^5, {\rm sup}}+\frac{112221208}{23625}\nonumber\\ %%%%%\frac{111851608}{23625}
&+&\frac{4225 \pi^2}{36}+\frac{794624 }{1575}\log (2p_\infty)\,.
\eea
We note in passing that the coefficient of $\pi^2$ in the combined result \eqref{eqpinf1} is significantly simpler than the one appearing in the original (not combined) expression.
Though only the $p_\infty \ln(2p_\infty)$ term is uniquely predicted by this result, it will be interesting to see whether the $\pi^2$ term in the latter result agrees with future 2SF-level $G^5$ scattering results. If it is the case, we are entitled to absorb $f_1^{G^5, {\rm sup}}$  in a redefinition of the rational part, $d_5^{\rm rat}$  of $\bar d_5^{\nu^2}$, say
\beq
\frac{8}{15}d_5^{\rm rat}{}'=\frac{8}{15}d_5^{\rm rat}+f_1^{G^5, {\rm sup}}\,.
\eeq
At the $O(p_\infty^2)$ level (5.5PNabs), the corresponding term in $f_b^{G^5\, {\rm 2SF}}(\gamma)$ reads
\beq
\label{pinfsquare}
C_{b2}=\frac{67876972}{363825} + f_2^{G^5, {\rm sup}} + \frac{23783}{3360} \pi^2 \,. %%%\frac{7619536}{40425} 
\eeq

As in the discussion of the $O(p_\infty^1)$ term, we expect our undetermined parameter $f_2^{G^5, {\rm sup}}$ to be, similarly to the presently undetermined rational (nonlocal-related) contribution to $\bar d_5^{\nu^2}$,  to be rational. In view of the fact that $f_2^{G^5, {\rm sup}}$ was an addition to the algebraically  simplest
solution  to the constraints i), ii), iii) above (see Eq. (12.51) in Ref. \cite{Bini:2022enm}) that have to be satisfied by $\Delta p_1^{\rm remain}$, it is even reasonable to expect that $f_2^{G^5, {\rm sup}}$ may simply vanish. An additional argument suggesting that $f_2^{G^5, {\rm sup}}$ might vanish is the fact that the direct impact of the $G^2$ radiation-reaction force on scattering proceeds via the $O(G^3)$ $\chi^{J_{\rm rad}}$ term whose square is $O(G^6)$ and not $O(G^5)$, see discussion in the Concluding Remarks below. Indeed, the latter fact indicates that there is  no ambiguity at $O(G^5)$ linked to radiation-reaction squared effects. 
In this case both the $\pi^2$ contribution and the rational contribution in Eq. \eqref{pinfsquare} should agree with future 2SF-level $G^5$ scattering results.

At $O(p_\infty^3)$ the contribution to $f_b^{G^5\,{\rm 2SF}}(\gamma)$ reads
\bea
C_{b3}&=&\frac{8}{35}q_{4,5}^{\nu^2}+f_3^{G^5, {\rm sup}}+\frac{64192021}{308700} \nonumber\\ %%%%\frac{2854813}{12348}\nonumber\\ 
&+&\frac{28637}{360} \pi ^2+\frac{841216}{2205} \ln(2p_\infty)\,.
\eea 

Again it will be interesting to compare this TF-based result with future 2SF $G^5$ results (contrary to the 5PNabs level parameter $\bar d_5^{\nu^2}$ the transcendental part of 6PN-level one $q_{4,5}^{\nu^2}$ has not been determined).

Let us finally emphasize that one of the defining characteristics of the TF approach is to point out the presence of the angular momentum loss in the radiation reacted impulse. It would be interesting to understand this occurrence within the WEFT formalism. Furthermore, when the $O(G^4)$ angular momentum loss is computed 
it will be interesting to compare the 2SF-level TF prediction %(general eq for f_b^{2SF}) 
to the corresponding 2SF  EFT impulse result.

\section{Concluding Remarks}

In the present work we summarized the results of the TF formalism concerning black hole scattering that were disseminated in the series of papers that developed the Tutti Frutti formalism (starting with Ref. \cite{Bini:2019nra} and including \cite{Bini:2020wpo,Bini:2020nsb,Bini:2020hmy,Bini:2020rzn}). 
We found a way to simplify the structure of the TF-obtained radiation-reacted impulse by rewriting it in the following form (see Eq. \eqref{deco_fin}):
\bea
\label{deco_fin2}
\Delta p_a^\mu  &=& \Delta p_a^{{\rm cons}\,\mu}+\Delta  p_{a }^{{\rm rr  lin} J_{\rm rad}\mu}\nonumber\\
&+&\Delta  p_{a }^{{\rm rr  lin} P_{\rm rad}\mu}+\Delta  p_{1}^{ \rm rr \, remain \, sup}{}^\mu\,.
\eea
Here $\Delta p_a^{{\rm cons}\,\mu}$ was defined in Eq. \eqref{eq2.1} above, $\Delta  p_{a }^{{\rm rr  lin} J_{\rm rad}\mu}$ was defined in Eq. \eqref{deltapradJ}, $\Delta  p_{a }^{{\rm rr  lin} P_{\rm rad}\mu}$ was defined in Eq. \eqref{deltaplinP} and $\Delta  p_{1}^{ \rm rr \, remain \, sup}{}^\mu$ was defined in Eq. \eqref{eqremainsup}.
This form is a 4-vectorial generalization of the original linear-response  formula of Ref. \cite{Bini:2012ji} in that it decomposes the total impulse $\Delta p_a^\mu  $ in: 1) a conservative contribution; 2) a radiative contribution linear in the c.m.-radiated angular momentum, $J_{\rm rad}$; 3)  a radiative contribution linear in the radiated four momentum $P_{\rm rad}^\mu$; and 4) a supplementary contribution which starts only at the $G^5$ order.
Both  $\Delta  p_{a }^{{\rm rr  lin} J_{\rm rad}\mu}$ and $\Delta  p_{a }^{{\rm rr  lin} P_{\rm rad}\mu}$ start at order $G^3$,
\bea
\label{ordersinG}
\Delta  p_{a }^{{\rm rr  lin} J_{\rm rad}\mu}&\sim & G^3+G^4+G^5+\ldots \nonumber\\
\Delta  p_{a }^{{\rm rr  lin} P_{\rm rad}\mu}&\sim & G^3+G^4+G^5+\ldots \,.
\eea
By definition of the TF formalism, each term in the  decomposition \eqref{deco_fin2} is polynomial in the two masses. In addition, defining
\beq
0=p_{a\, \rm out}^{{\rm cons}\, \mu}=m_a u_a^\mu +\Delta p_a^{{\rm cons}\,\mu}\,,
\eeq 
taking into account that $p_{a\, \rm out}^{{\rm cons}\, \mu}$ separately satisfies the mass-shell condition 
$(p_{a\, \rm out}^{{\rm cons}})^2=(p_{a\, \rm in}^{{\rm cons}})^2=-m_a^2$
and remembering  that $\Delta  p_{a }^{{\rm rr  lin} J_{\rm rad}\mu}$ and $\Delta  p_{a }^{{\rm rr  lin} P_{\rm rad}\mu}$ both start at order $G^3$, Eq. 
\eqref{ordersinG}, the mass-shell constraint on the total, radiation-reacted impulse reads
\bea
0&=&(p_{a\,\rm out}^{{\rm cons}}+\Delta  p_{a }^{{\rm rr  lin} J_{\rm rad}}+\Delta  p_{a }^{{\rm rr  lin} P_{\rm rad}})-p_{a\, \rm in}^2\nonumber\\
&=&2
p_{a\,\rm out}^{{\rm cons}}\cdot \Delta  p_{a }^{{\rm rr  lin} J_{\rm rad}}+2
p_{a\,\rm out}^{{\rm cons}}\cdot \Delta  p_{a }^{{\rm rr  lin} P_{\rm rad}}\nonumber\\
&+&O(G^6)\,. 
\eea 
Each term on right-hand-side of the latter equation (which starts at order $G^3$) vanishes separately.
We note that if one extends  the definition of the $J_{\rm rad}$ part of the impulse by defining it as
being the all-order $G$-expansion of, for example,  
$\Delta p_1^{{\rm cons}\mu}(\chi^{\rm cons}(J-\frac12 J_{\rm rad}))$,  
the so defined radiation-reacted nonlinear-in-$J_{\rm rad}$ impulse, $\Delta p_1^{{\rm nonlin},J_{\rm rad}}$, would {\it exactly} satisfy the mass-shell condition.
Any such definition involving   nonlinear-in-$J_{\rm rad}$ extensions of $J$ differs anyway from $\Delta p_1^{{\rm nonlin},J_{\rm rad}}$ at $O(G^6)$.

At order $G^4$ we compared the TF predicted result \eqref{deco_fin2} to the state-of-the-art post Minkowskian results. 
We highlighted the agreement between the results of Ref. \cite{Dlapa:2022lmu} and  two structural features predicted by   the TF formalism:  i) the simple proportionality  relation between the \lq\lq 2rad" dissipative component along $b$ computed in \cite{Dlapa:2022lmu} and the projection of the $O(G^4)$ radiated four momentum along the vector $e_x^\mu$:
\beq
c_{\rm 1b,2rad}^{\rm 4diss}=\frac{m_1}{m_2-m_2}P_{xG^4}^{\rm rad}\,;
\eeq
ii) the possibility to extract the exact value of the $O(G^3)$ radiated angular momentum from the \lq\lq 1rad" dissipative component along $b$ computed in \cite{Dlapa:2022lmu}. As exhibited in Eqs. \eqref{eqJ31} and \eqref{eqJ32}, the so extracted value of $\hat J_3$ coincides with the results obtained in Ref. \cite{Manohar:2022dea}. In addition we recalled that 
the PN-accurate expansion of the radiation reacted impulse which was computed in TF at the absolute 6PN accuracy agreed with the PN expansion of the   exact $G^4$ result of \cite{Dlapa:2022lmu}. 

At order $G^5$, we compared the TF predicted result \eqref{decoatG5} to the state-of-the-art post Minkowskian results~\cite{Driesse:2024feo}, which are currently limited to the 1SF level in the sense of the mass decomposition \eqref{plefka_eqs} and \eqref{plefka_eqs2} (i.e., terms with first-order in either $m_1/m_2$ or $m_2/m_1$). 
The TF predictions for the structure of the 1SF $G^5$ impulse were given in Eqs. \eqref{f5TF1} and \eqref{f5TF2}, where we recall that the notation \lq\lq K" denotes terms known from previous orders. The full expressions of the TF-predicted  impulse coefficients are given in a Supplemental Material file. 
We highlighted the presence of the  $\nu^2$-rescaled $O(G^4)$ angular momentum loss $\hat J_4^0(\gamma)$, as defined in the third of Eqs. \eqref{hatJ_defs} (see Eq. \eqref{J40combined} for its PN evaluation).
When the WEFT formalism will compute the exact value of $\hat J_4^0(\gamma)$ one will be able to read-off the exact value of $\chi_5^{{\rm cons}1}$ from \cite{Driesse:2024feo}.
At the time being we know  from the TF formalism high-PN accuracy values for all the ingredients entering the TF-predicted impulse, see Eqs. \eqref{J40combined}, \eqref{P1m2G5}, \eqref{P5bgamma}, \eqref{P1p250}, \eqref{P1p250bis} displayed in Section VI.
In particular, we highlight again that a subtle nonlocal 5.5PN tail-squared contribution to $\tilde \chi_5^{\rm cons}$ enters $f_b^{G^5 {\rm 1SF}} $ and agrees with the corresponding recent results of \cite{Driesse:2024feo}. 
We presented  in Eqs.  \eqref{eqtailoftail0} and \eqref{eqtailoftail}  the all-multipole generalization of the quadrupolar-level action
of  Refs.  \cite{Damour:2015isa,Bini:2020wpo}  describing the $O(G^5)$ nonlocal-in-time  conservative dynamics, mediated by time-symmetrized tail-squared interactions. We leave to future works a corresponding PN-extension of the tail-of-tail contributions to scattering, Eq. (5.10) in Ref. \cite{Bini:2020wpo} where it was first derived (see Eq. \eqref{chi5p5} here).
It would be interesting to have an independent derivation of the nonlocal tail-of-tail action within the NRGR approach. 

We also found agreement between TF and WEFT at the next, 6PN level. 
This agreement was obtained by combining  TF results with a recent, partial, computation of $\hat J_4^0$ 
\cite{Heissenberg:2025ocy}.  

The TF-predicted PN expansion of three more independent functions of $\gamma$ related to radiated linear momentum, $\hat P_{1-2}^{G^5}$, $\hat P_b$, $\hat P_{102}^{5,0}$
have been found to be in perfect agreement with corresponding results in Ref. \cite{Driesse:2024feo}. 

We highlighted in Eqs. \eqref{Pb1p21m2G^5} 
 the extent to which the linear momentum balance, $\Delta p_1^\mu +\Delta p_2^\mu +P_{\rm rad}^\mu=0$ determines the expansion coefficients of the individual impulses $\Delta p_1$ and $\Delta p_2$.
When combining these constraints with the  mass-shell constraint, Eq. \eqref{masshelfine}, the main conclusion is that, at order $G^5$, all the separate impulse components are determined except for the average  
$\frac12 (f_b^{G^5,{\rm 1SF}}+f_b^{G^5,{\overline  {\rm 1SF}}})$  and $f_b^{G^5,{\rm 2SF}}$.
Concerning the 1SF level,  only the difference $ f_b^{G^5,{\rm 1SF}}-f_b^{G^5,{\overline  {\rm 1SF}}} $ is determined from the $b$ component of $P_{\rm rad}^\mu$ (together with known lower-order results). At the 1SF level, the resulting ambiguity in the separate determination of $ f_b^{G^5,{\rm 1SF}}$ and $f_b^{G^5,{\overline  {\rm 1SF}}} $ is completely lifted in the TF formalism which expresses both  $ f_b^{G^5,{\rm 1SF}}$ and $f_b^{G^5,{\overline  {\rm 1SF}}} $ in terms of the three TF-meaningful observables: $\tilde \chi^{1\,\rm cons}_5$, $\hat J_4^0$ and $P_b^5$. We note in particular that the fact that  $\hat P_b^5$ contributes only to $f_b^{G^5,{\overline  {\rm 1SF}}} $ and not to $ f_b^{G^5,{\rm 1SF}}$ is related to the constraint stated in \cite{Bini:2022enm} that terms in $\Delta p_1$ which are nonlinear in the radiation reaction force acting on body 1, should contain a factor $m_1^3$.
[At $G^4$ the same argument allowed the TF formalism to predict the full structure of $\Delta p_1^{\rm rr}$ in terms of $P_{\rm rad}^\mu$ and $J_{\rm rad}^\mu$.]

However, at the 2SF level there exists a remaining ambiguity in the TF determination of $\Delta p_{1b}^{\rm 2SF}$, as displayed in the first Eq. \eqref{f5TF3}. The TF formalism expresses $f_b^{G^5{\rm 2SF}}$ in terms of six ingredients, namely $f_b^{G^5}$, $\hat P_b^5$, $\tilde \chi_5^{1\rm cons}$, $\tilde \chi_5^{2\rm cons}$, $\hat J_4^0$ and  $\hat J_4^1$. Among these ingredients, the last five of them are TF-predicted at the current 6PN accuracy as displayed in Section VII.  

On the other hand, the function $f_b^{G^5\,{\rm remain}}(\gamma)$ entering the supplementary impulse contribution $\Delta p_1^{\rm rr\, remain\, sup}$, Eq. \eqref{eqremainsup}, is only constrained by the TF formalism to start at the 5PN order. 
As a consequence, the TF predictions for the value of $f_b^{G^5\, {\rm remain}}(\gamma)$ is only fully predicted at the 4.5PNabs accuracy, as displayed in Eq. \eqref{fbfinal}.
Beyond this order, as discussed in Section VII, the TF formalism  can however predict the value of the logarithmic coefficient at 5PN order.
Moreover, if one assumes that the supplementary contribution has rational PN-expansion coefficients, Eq. \eqref{eqpinf1} predicts the $\pi^2$ term in the 5PN contribution $f_{b, {\rm TF}}^{G^5, {\rm 2SF}}(\gamma)$.

\section*{Acknowledgements}

The present research was partially supported by the
2021 Balzan Prize for Gravitation: Physical and Astrophysical
Aspects, awarded to T. Damour.  We thank Andrea Geralico and Julio Parra-Martinez for informative discussions.
D.B. acknowledges 
sponsorship of the Italian Gruppo Nazionale per
la Fisica Matematica (GNFM) of the Istituto Nazionale
di Alta Matematica (INDAM), as well as the hospitality
and the highly stimulating environment of the Institut
des Hautes Etudes Scientifiques.

\appendix

\section{Notation and useful relations}
\label{app_notation}

We collect here some definitions and relations which are often used. 
We use the mostly positive signature.

The masses of the two bodies are denoted as $m_1$ and $m_2$, with the convention $m_1 \leq m_2$.
The symmetric mass ratio $\nu\equiv (m_1 m_2)/(m_1+m_2)^2$ is the ratio of the reduced mass $\mu\equiv (m_1 m_2)/(m_1+m_2)$ to the total mass $M=m_1+m_2$.

The  4-momenta in the asymptotic future ($t\to +\infty $, outcoming) are denoted by $p_{a}^{\rm out} = m_a u_{a}^{\rm out}$, with $a=1,2$, and the asymptotic energies by $E_a^{\rm out}$.
To simplify notation, in the asymptotic past  ($t\to -\infty $, incoming) we will often omit the corresponding \lq\lq in" label for all these quantities.
We often work within the incoming c.m. frame of the system, with time axis
\beq
U \equiv \frac{ p_1  + p_2 }{ |p_1 +  p_2 |}
=\frac{m_1u_1 +m_2u_2 }{E}\,,
\eeq
where $E=E_1+E_2$ is the incoming c.m. energy, which is related to the incoming value of the momentum by
\beq
P_{\rm c.m.}= \frac{m_1m_2}{E}\sqrt{\gamma^2-1}= \frac{m_1m_2 }{E}\pinf
\,,
\eeq
with
\beq
\g \equiv  -u_1 \cdot u_2  = -  \frac{p_1 \cdot p_2 }{m_1 m_2}\,,\qquad
\pinf \equiv \sqrt{\g^2-1}\,,
\eeq
so that $E P_{\rm c.m.} = m_1 m_2 \pinf$.
The total incoming energy can also be written as
\beq
\frac{E}{Mc^2} \equiv h(\g, \nu)=\sqrt{1+2\nu (\g-1)}
\,,
\eeq 
implying that $P_{\rm c.m.} =\mu {\pinf}/{h}$ and
\beq
\frac{G E}{b} \equiv \frac{GM h}{b} = \frac{\pinf}{j}\,,
\eeq
where $b$ is the impact parameter and
\beq 
 j\equiv \frac{c J_{\rm c.m.} }{G m_1 m_2}=\frac{c J_{\rm c.m.}}{G M \mu}\,,
 \eeq
is a dimensionless rescaled version of the total center-of-mass angular momentum $J_{\rm c.m.}$ at $t\to -\infty$.
The individual incoming energies (see Eqs. (A9) of Ref. \cite{Bini:2021gat}) read
\beq
\frac{E_1}{m_1}=\frac{m_1+\g m_2 }{E}\,,\qquad
\frac{E_2}{m_2}=\frac{m_2+ \g m_1}{E}\,,
\eeq
so that, for example, 
\beq
\label{en12}
\frac{ m_1 E_2 + m_2 E_1}{E}=m_1m_2 (m_1+m_2)\frac{(\gamma+1)}{E^2}\,.
\eeq

The vectorial  impact parameter (orthogonal to $ u_1$ and  $ u_2$)
 ${\mathbf b}_{12}= {\mathbf b}_{1}- {\mathbf b}_{2}=b \hat {\mathbf b}_{12}$ together with the conservative scattering angle $\chi_{\rm cons}$ enters the definition of the Cartesianlike basis vectors ${\mathbf e}_x$ and ${\mathbf e}_y$ as follows (see Eq. (3.49) of Ref. \cite{Bini:2021gat}):
\bea
\label{exydef}
{\mathbf e}_x&=& \cos \frac{\chi_{\rm cons}}{2} \hat {\mathbf b} + \sin \frac{\chi_{\rm cons}}{2}  {\mathbf n}  \,,\nonumber\\
{\mathbf e}_y&=& -\sin \frac{\chi_{\rm cons}}{2} \hat {\mathbf b} + \cos \frac{\chi_{\rm cons}}{2}  {\mathbf n}\,,
\eea
where ${\mathbf n}$ is the direction of the incoming momenta (orthogonal to the c.m. four velocity, $n\cdot U=0$), 
\beq
{\mathbf n}=\frac{m_1 m_2}{P_{\rm c.m.} E}\left(\frac{E_2}{m_2}u_1 -\frac{E_1}{m_1}u_2\right)=\frac{(u_{2}\wedge u_{1})\cdot U}{\sqrt{\gamma^2-1}}\,.
\eeq
Here the wedge product of two vectors $A$ and $B$ is standardly defined as $A\wedge B= A\otimes B-B\otimes A $.

Boldface vectors denote spatial vectors in the c.m. frame with time axis $U$: $p_a=m_a u_a=E_a U  +{\mathbf p}_a$ (where ${\mathbf p}_a$ is orthogonal to $U$,
and ${\mathbf p}_1=-{\mathbf p}_2={\mathbf P}_{\rm c.m.}$).
When describing the conservative scattering it is useful to introduce the c.m. direction of the (conservative) outgoing
momenta, ${\mathbf n}_{\rm out}^{\rm cons}$, as well as its associated orthogonal direction $\hat {\mathbf B}$, namely
\bea
\hat {\mathbf B}&=& \cos(\chi_{\rm cons})\hat {\mathbf b} +\sin (\chi_{\rm cons}){\mathbf n}  \,,\nonumber\\
{\mathbf n}_{\rm out}^{\rm cons}&=&-\sin(\chi_{\rm cons})\hat {\mathbf b} +\cos (\chi_{\rm cons}){\mathbf n} 
\,.
\eea
Note the relation
\beq
\hat {\mathbf B}= -\frac{d}{d \chi_{\rm cons}}{\mathbf n}_{\rm out}^{\rm cons}\,.
\eeq
The dyad $ (\hat {\mathbf B}, {\mathbf n}_{\rm out}^{\rm cons},) $ differs from the incoming dyad
$(\hat {\mathbf b}, {\mathbf n})$ by a rotation of angle $\chi_{\rm cons}$. The dyad $({\mathbf e}_x ,{\mathbf e}_y)$
is midway between the latter two dyads, being obtained from the incoming dyad by a rotation of angle  $\frac12 \chi_{\rm cons}$.

\section{PN-expanded forms of various quantities}

\begin{table*}  
\caption{\label{PNvarious} PN-expanded forms of the various quantities at $O(G^3)$, $O(G^4)$ and $O(G^5)$.
}
\begin{ruledtabular}
\begin{tabular}{ll}
$\hat P_{1+2}^3$ 
& $\pi \left[ \frac{37  }{30}p_\infty+\frac{839  }{1680}p_\infty^3+\frac{2699 }{2016} p_\infty^5-\frac{1531643 }{1182720}  p_\infty^7
+\frac{70348799}{61501440}  p_\infty^9-\frac{30685679  }{30750720}p_\infty^{11}+O\left(p_\infty^{13}\right)\right]$\\
$\hat J_2$ &
$\frac{16 }{5}p_\infty^3+\frac{176 }{35}p_\infty^5-\frac{608}{315} p_\infty^7+\frac{3712 }{3465}p_\infty^9-\frac{10496 }{15015}p_\infty^{11}+O\left(p_\infty^{13}\right)$\\
$\hat J_3$ 
&$ \pi \left[\frac{28}{5}p_\infty^2+\frac{739}{84} p_\infty^4-\frac{5777}{2520}p_\infty^6+\frac{115769}{126720} p_\infty^8-\frac{16548173 
  }{23063040} p_\infty^{10}+\frac{3912731}{6589440} p_\infty^{12}
+O\left(p_\infty^{13}\right)\right]$\\
$\hat J_4^0$ &$ \frac{176}{5}p_\infty+\frac{8144}{105} p_\infty^3+\frac{448}{5} p_\infty^4
-\frac{93664}{1575} p_\infty^5
+\frac{1184}{21}p_\infty^6
-\frac{4955072}{121275}p_\infty^7
-\frac{13736}{315}  p_\infty^8+O(p_\infty^9)$\\
$\hat J_4^1$&$  -\frac{208}{15}  p_\infty^3 
+ \frac{988}{63}  p_\infty^5
-\frac{13312}{525} p_\infty^6
+ \frac{5458}{1575}  p_\infty^7
-\frac{208}{225}  p_\infty^8+O(p_\infty^9)$\\
$\tilde E_4^0(\gamma)$& $\frac{1568}{45} p_\infty^3+\frac{18608}{525} p_\infty^5+\frac{3136}{45} p_\infty^6+\frac{220348}{11025} p_\infty^7+\frac{1216
  }{105} p_\infty^8-\frac{151854}{13475} p_\infty^9+\frac{117248}{1575} p_\infty^{10}-\frac{405087523
   }{9909900}p_\infty^{11} -\frac{224512}{3465} p_\infty^{12}+O(p_\infty^{13})$\\
$\tilde E_4^1(\gamma)$ &$  -\frac{352}{45} p_\infty^5+\frac{1736
   }{225}p_\infty^7-\frac{704}{45} p_\infty^8+\frac{8068}{4725} p_\infty^9-\frac{1808}{225} p_\infty^{10}+\frac{1967239
   }{103950}p_\infty^{11}-\frac{47176 }{4725}p_\infty^{12}
+O(p_\infty^{13})$\\
$\hat P_b^4$ 
&$-\pi \left[\frac{37}{30}+\frac{1661}{560}p_\infty^2+\frac{1491}{400}p_\infty^3+\frac{23563}{10080}p_\infty^4-\frac{26757}{5600} p_\infty^5
+\frac{700793}{506880}  p_\infty^6+\frac{4046561}{451584} p_\infty^7-\frac{8085643 }{5591040}p_\infty^8-\frac{628523873 }{41395200} p_\infty^9\right.$\\
&$\left.+\frac{23609 }{18304}p_\infty^{10}+\frac{3709460060017}{154983628800}  p_\infty^{11}-\frac{56351596657}{50185175040}  p_\infty^{12}+O(p_\infty^{13})\right]$\\
$\hat P_{1+2}^4$ 
&$\frac{784}{45 p_\infty}+\frac{2168}{175}p_\infty +\frac{1568}{45} p_\infty^2+\frac{98666 }{11025}
-\frac{512}{105} p_\infty^4-\frac{2702747}{363825}p_\infty^5+\frac{62392 }{1575}p_\infty^6 -\frac{2178650927 }{138738600}p_\infty^7
-\frac{2335876}{51975}p_\infty^8$\\
&$+\frac{11696625509}{327927600} p_\infty^9+\frac{140685967 }{3153150}p_\infty^{10}-\frac{128153894712407
  }{2207608603200} p_\infty^{11} 
-\frac{266660131 }{6306300}p_\infty^{12}+O(p_\infty^{13})$\\
$\hat P_{1-2}^4$
&$  \frac{176}{45 p_\infty}-\frac{72 }{25}p_\infty+\frac{352}{45} p_\infty^2-\frac{9746}{4725} p_\infty^3
+\frac{448}{75} p_\infty^4-\frac{484019}{51975} p_\infty^5+\frac{26024}{4725} p_\infty^6+\frac{157355621 }{46246200}p_\infty^7
-\frac{79612}{14553} p_\infty^8 $\\
&$ -\frac{1100898451 }{1545944400}p_\infty^9+\frac{46101421 }{9459450}p_\infty^{10}-\frac{87328389299 }{43286443200}p_\infty^{11}-\frac{629179 }{147420}p_\infty^{12}+O(p_\infty^{13}) $
\end{tabular}
\end{ruledtabular}
\end{table*}

We list in Table \ref{Schwchi} the exact expressions for the scattering angle in the Schwarzschild spacetime at various PM orders.

\begin{table*}  
\caption{\label{Schwchi} Exact expressions for the scattering angle in the Schwarzschild spacetime.
}
\begin{ruledtabular}
\begin{tabular}{ll}
$\chi_1^{\rm Schw}(p_\infty)$&$ \frac{1}{p_\infty}+2 p_\infty  $\\
$\chi_2^{\rm Schw}(p_\infty)$&$\pi \left(\frac32 + \frac{15}{8}  p_\infty^2 \right)$\\
$\chi_3^{\rm Schw}(p_\infty)$&$-\frac{1}{ 3 p_\infty^3}+\frac{4}{p_\infty} +24 p_\infty + \frac{64}{3}  p_\infty^3$\\
$\chi_4^{\rm Schw}(p_\infty)$&$ \pi \left(\frac{105}{8}+  \frac{315}{8} p_\infty^2+\frac{3465}{128} p_\infty^4\right)$\\
$\chi_5^{\rm Schw}(p_\infty)$&$\frac{1}{5 p_\infty^5}-\frac{2}{p_\infty^3}+\frac{ 32 }{p_\infty} +320p_\infty + 640 p_\infty^3 
+\frac{1792}{5} p_\infty^5$\\
$\chi_6^{\rm Schw}(p_\infty)$&$ \pi\left(
\frac{1155}{8}+\frac{45045}{64} p_\infty^2  + \frac{135135}{128} p_\infty^4
+\frac{255255}{512}p_\infty^6\right)$\\
$\chi_7^{\rm Schw}(p_\infty)$&$
-\frac{1}{7 p_\infty^7}+\frac{8}{5 p_\infty^5}  -  \frac{16}{p_\infty^3}  + \frac{320}{p_\infty} +4480 p_\infty+14336 p_\infty^3 +\frac{86016}{5}p_\infty^5 + \frac{49152}{7}p_\infty^7$\\
\end{tabular}
\end{ruledtabular}
\end{table*}

\subsection{$P_b$ and $P_n$ vs $P_{1\pm 2}$ at all orders in $G$}

Starting from the relations (valid at all orders in $G$)
\bea
2P_{1+2}&=& \frac{E^{\rm rad}}{h}+\frac{P_n}{\sqrt{\gamma^2-1}}\left(\frac{E_2}{m_2}-\frac{E_1}{m_1}\right)\,, \nonumber\\
2P_{1-2}&=& \frac{(m_1-m_2)}{M h}E^{\rm rad}+\frac{P_n}{\sqrt{\gamma^2-1}}\left(\frac{E_2}{m_2}+\frac{E_1}{m_1}\right) \,,
\nonumber\\
\eea
where
\bea
\frac{E_2}{m_2}-\frac{E_1}{m_1}&=& \frac{(\gamma-1) (m_1 - m_2)}{Mh}\,, \nonumber\\
\frac{E_2}{m_2}+\frac{E_1}{m_1}&=& \frac{(\gamma+1) (m_1 + m_2)}{Mh} \,,
\eea
we find 
\bea
P_{1+2}&=& \frac{1}{2h}\left[E^{\rm rad}+\frac{(m_1 - m_2)}{M} \sqrt{\frac{\gamma-1}{\gamma+1}}P_n \right]\,, \nonumber\\
P_{1-2}&=& \frac{1}{2h}\left[\frac{(m_1-m_2)}{M}E^{\rm rad}+ \sqrt{\frac{\gamma+1}{\gamma-1}}P_n  \right]\,.
\eea

\subsection{$P_b,P_n$ vs $P_x,P_y$ and their expressions  in powers of $1/j$ and $1/b$}

The PM expansions of  $P_b$ and $P_n$ in powers of $GM/b$ read
\bea
P_b &=& (m_2-m_1)\nu^2  \hat P_b^{(b)}\,,\qquad P_n= (m_2-m_1)\nu^2 \hat P_n^{(b)}\,,\qquad
\eea
where
\label{Pnpb}
\bea
\hat P_b^{(b)}&=& \hat P_b^{3,(b)} \left(\frac{GM}{b} \right)^3+ \hat P_b^{4,(b)} \left(\frac{GM}{b} \right)^4\nonumber\\
&+&\hat P_b^{5,(b)} \left(\frac{GM}{b} \right)^5+O(G^6)\,,\nonumber\\ 
\hat P_n^{(b)}&=& \hat P_n^{3,(b)} \left(\frac{GM}{b} \right)^3 +\hat P_n^{4,(b)}\left(\frac{GM}{b} \right)^4\nonumber\\ 
&+& \hat P_n^{5,(b)}\left(\frac{GM}{b} \right)^5+O(G^6)\,.
\eea
Let us also recall the relations between $j$-expanded quantities (referred to the frame $({\mathbf e}_x,{\mathbf e}_y)$) and $b$-expanded quantities  (referred to the frame $(\hat {\mathbf b},{\mathbf n})$), namely
\bea
\hat P_b^{3,(b)}&=& 0\,,\nonumber\\ 
\hat P_b^{4,(b)} &=& -\frac{h^4}{(\gamma^2-1)^2} (-\check P_x^4+\chi_1^{\rm cons} \check P_y^3)\,,\nonumber\\ 
\hat P_b^{5,(b)} &=& -\frac{h^5}{(\gamma^2-1)^{5/2}} (\chi_1^{\rm cons} \check P_y^4+\chi_2^{\rm cons} \check P_y^3  -\check P_x^5)\,, \nonumber\\
\hat P_n^{3,(b)} &=& \check P_y^3\,,\nonumber\\
\hat P_n^{4,(b)} &=& \check P_y^4\,,\nonumber\\
\hat P_n^{5,(b)}&=& \check P_y^5-\frac12  (\chi_1^{\rm cons})^2 \check P_y^3+\chi_1^{\rm cons} \check P_x^4\,,
\eea
where $\check P_x^n$ and $\check P_y^n$ and $\hat P_b^{n,(b)}$ are defined such that 
\bea
P_i=(m_2-m_1)\nu^2 \sum_{n\ge 3}\frac{\check P_{i}^n}{j^n}\,,\qquad i=x,y\nonumber\\
P_b=(m_2-m_1)\nu^2 \sum_{n\ge 4}\hat P_b^{n,(b)}\left(\frac{GM}{b}\right)^n\,.
\eea

\begin{table*}  
\caption{\label{Schwchi} First terms in the PN expansion of $\hat P_b^{4,(b)}$, $\hat P_b^{5,(b)}$, $\hat P_n^{3,(b)}$, $\hat P_n^{4,(b)}$ and $\hat P_n^{5,(b)}$.
}
\begin{ruledtabular}
\begin{tabular}{ll}
$\hat P_b^{4,(b)}$ &$\pi\left[ -\frac{37}{30} -\frac{1661}{560} p_\infty^2 -\frac{1491}{400} p_\infty^3 -\frac{23563}{10080} p_\infty^4 
+\frac{26757}{5600} p_\infty^5 -\frac{700793}{506880} p_\infty^6-\frac{238033}{26564} p_\infty^7+O(p_\infty^8)\right]$\\
$\hat P_b^{5,(b)}$&$-\frac{64}{ 3 p_\infty^2}-\frac{27392}{525}-\frac{37}{20}\pi^2-\frac{30208}{225} p_\infty
+\left(-\frac{856768}{33075}-\frac{3429}{1120}\pi^2\right) p_\infty^2
+\frac{462592}{7875} p_\infty^3  \left(-\frac{7915}{2688}\pi^2-\frac{74417152}{363825} \right) p_\infty^4-\frac{99136}{2625}p_\infty^5+O(p_\infty^6)$\\
\hline
$\hat P_n^{3,(b)}$ & $ \pi\left[ \frac{37}{30} p_\infty^2 
+\left(-\frac{37}{60}\nu +\frac{839}{1680} \right) p_\infty^4
+\left(-\frac{107}{1120}\nu+\frac{37}{80} \nu^2+\frac{2699}{2016} \right) p_\infty^6 +\left(-\frac{27581}{40320}\nu-\frac{197}{4480}\nu^2-\frac{37}{96}\nu^3-\frac{1531643}{1182720} \right) p_\infty^8+O(p_\infty^{10})\right]$\\
$\hat P_n^{4,(b)}$ & $ \frac{64}{3}+\left(-\frac{32}{3}\nu+\frac{1664}{175} \right) p_\infty^2
+\frac{128}{3} p_\infty^3
+\left(8\nu^2-\frac{1096}{525}\nu+\frac{227776}{33075} \right) p_\infty^4
+\left(-\frac{64}{3}\nu+\frac{192}{175}\right) p_\infty^5 $\\
&$+\left(-\frac{118676}{33075}\nu-\frac{76}{175}\nu^2-\frac{20}{3}\nu^3-\frac{1218176}{72765}\right) p_\infty^6
+\left(\frac{8528}{189} + \frac{2512}{525} \nu + 16 \nu^2\right)p_\infty^7+O(p_\infty^8)$\\
$\hat P_n^{5,(b)}$& $\pi\left[ \frac{53}{ 3p_\infty^2}+\frac{60563}{5040} -\frac{161}{12}\nu 
+\left(-\frac{1491}{400}+\frac{1509}{140}\pi^2\right) p_\infty
+\left(-\frac{1298161}{40320} -\frac{7969}{2520}\nu +\frac{107}{12}\nu^2 \right) p_\infty^2 \right.$\\
&$+\left(\frac{2678867}{16800}-\frac{218867}{67200}\nu -\frac{75661}{4480}\pi^2+\frac{491447}{67200}\nu -\frac{3957}{560}\pi^2\nu \right) p_\infty^3$\\
&$+\left(\frac{2435}{8064}\nu^2-\frac{695}{96} \nu^3 -\frac{41053}{2450}\ln(p_\infty)+\frac{32135}{768}\nu +\frac{194769745661}{4346496000} -\frac{4059}{1280}\pi^2\nu+\frac{503}{70}\pi^2+\frac{41053}{2450}\ln(2)\right) p_\infty^4$\\
&$+\left. \left( -\frac{2678867}{16800} +  \frac{22984309}{67200} \nu - 
 \frac{7981}{3200} \nu^2  + \frac{75661}{4480}\pi^2  -  \frac{41863}{1280 } \nu \pi^2 + \frac{2733}{560} \nu^2\pi^2+ \frac12 \nu E_{5,9}^1\right)p_\infty^5+O(p_\infty^6)\right]$\\
\end{tabular}
\end{ruledtabular}
\end{table*}

\section{Polynomiality structure of various quantities}
\label{appendixex}

\begin{widetext}
From the expansions 
\bea
\sin \left(\frac{\chi^{\rm cons}}{2}\right)&=&\frac{GMh}{b}  \left[ \frac{\chi^{\rm cons}_1}{p_\infty}   
+ \frac{\tilde \chi^{\rm cons}_2}{p_\infty^2}  \left(\frac{GM}{b} \right)^1 
+ \left(\frac{\tilde \chi^{\rm cons}_3}{p_\infty^3} - \frac{(\chi^{\rm cons}_1)^3 h^2}{6 p_\infty^3}\right)
\left(\frac{GM}{b} \right)^2 +\ldots\right]\,,\nonumber\\
\cos \left(\frac{\chi^{\rm cons}}{2}\right)&=& 1  - \frac{(\chi^{\rm cons}_1)^2 h^2}{2 p_\infty^2} \left(\frac{GM}{b} \right)^2
-\frac{\tilde \chi^{\rm cons}_2 h^2\chi^{\rm cons}_1}{ p_\infty^3} \left(\frac{GM}{b} \right)^3+\ldots \,,
\eea
\end{widetext}
we recognize that $
E^{-1}\sin \left(\frac{\chi^{\rm cons}}{2}\right)=\frac{1}{Mh}\sin \left(\frac{\chi^{\rm cons}}{2}\right)
$ and $\cos \left(\frac{\chi^{\rm cons}}{2}\right)$
are two polynomial function of $m_1$ and $m_2$.
For the vector 
\bea
n&=&\frac{1}{\sqrt{\gamma^2-1}}\left(\frac{E_2}{m_2}u_1-\frac{E_1}{m_1}u_2 \right)\nonumber\\
&=&\frac{\sqrt{\gamma^2-1}}{E}(m_2 \check u_1 -m_1 \check u_2)\,, 
\eea
we see that the product $E n$ is linear in the masses $m_1$ and $m_2$.
Consequently, recalling
\bea
e_x &=& \cos \frac{\chi^{\rm cons}}{2} \hat b+\sin \frac{\chi^{\rm cons}}{2} n\nonumber\\
&=&\cos \frac{\chi^{\rm cons}}{2} \hat b+E^{-1}\sin \frac{\chi^{\rm cons}}{2} (E n)\nonumber\\
E e_y &=& -E \sin \frac{\chi^{\rm cons}}{2} \hat b+\cos \frac{\chi^{\rm cons}}{2} E n\nonumber\\
&=&  -E^2 \frac{\sin \frac{\chi^{\rm cons}}{2}}{E} \hat b+\cos \frac{\chi^{\rm cons}}{2} E n \,,
\eea
where $E^2=m_1^2 +m_2^2+2\gamma m_1m_2$, 
we have that $e_x^\mu$ and $E e_y^\mu$ are both polynomial in masses.

\begin{table*}  
\caption{\label{allfG5} PNexpansion of all the structure functions at $O(G^5)$.
}
\begin{ruledtabular}
\begin{tabular}{ll}
$f_{\check u_2}^{G^5\, {\rm 2SF}}$&$\pi \left[-\frac{18}{p_\infty^6}+\frac{114}{p_\infty^4}+\frac{60}{p_\infty^3}+\left(\frac{3409}{4}+\frac{123 \pi ^2}{64}\right)\frac{1}{p_\infty^2} 
+\frac{21787}{140 p_\infty}-\frac{148}{5}\log ( \frac{p_\infty}{2})+\frac{854981 \pi ^2}{20480}+\frac{115753}{96}+\frac{122533}{720} p_\infty\right.$\\
&$+ \left(-\frac{5501}{70} \log (\frac{p_\infty}{2} )-\frac{965609 \pi ^2}{40960}+\frac{24412151}{33600}\right)p_\infty^2+\left(-\frac{10593}{175}  \log (\frac{p_\infty}{2} )+\frac{8613 \pi ^2}{640}+\frac{69397802563}{155232000}\right)p_\infty^3$\\
&$\left.+\left( E^1_{5,9}-\frac{60521}{840} \log (\frac{p_\infty}{2} )-\frac{3109120219 \pi ^2}{58720256}+\frac{14402621}{39200}\right)p_\infty^4 +O(p_\infty^5)\right]$\\
$f_{\check u_1}^{G^5\, {\rm 2SF}}$& $\pi \left[
\frac{18}{p_\infty^6}-\frac{114}{p_\infty^4}-\frac{56}{5 p_\infty^3}+\left(-\frac{3409}{4}-\frac{123 \pi ^2}{64}\right)\frac{1}{p_\infty^2}
-\frac{15007}{210 p_\infty}-\frac{49345 \pi ^2}{4096}-\frac{115753 }{96}+\frac{148}{5}\log (\frac{p_\infty}{2} )-\frac{235867}{1260}  p_\infty\right.$\\
&$\left.+\left(\frac{5501}{70} \log (\frac{p_\infty}{2} )-\frac{227435 \pi ^2}{8192}-\frac{920693}{4800}\right)p_\infty^2- \frac{11646937 }{63360}p_\infty^3+\left(\frac{60521}{840} \log (\frac{p_\infty}{2} )-\frac{219729011 \pi ^2}{8388608}+ \frac{20028103}{58800}\right)p_\infty^4 +O(p_\infty^5)\right]$\\
\hline
$f_{b}^{G^5\, {\rm 1SF}}$&$ -\frac{8}{p_\infty^9}+\frac{44}{p_\infty^7}+\frac{329+36 \pi ^2}{p_\infty^5}+\frac{416}{45 p_\infty^4}+\left(\frac{889}{6}+\frac{643 \pi^2}{4}\right)\frac{1}{p_\infty^3}+\left(\frac{203264}{1575}-\frac{47 \pi ^2}{5}\right)\frac{1}{p_\infty^2}$\\
&$+\left(\frac{12544}{45} \log (2p_\infty )+\frac{14695 \pi ^2}{72}+\frac{575123}{2160} \right)\frac{1}{p_\infty}+\frac{116992}{1225}-\frac{5697 \pi ^2}{280}+\left(\frac{148864}{525}\log (2p_\infty )-\frac{33019 \pi^2}{480}+\frac{534248951}{252000}\right)p_\infty$\\
&$+\left(\frac{3407 \pi ^2}{672}-\frac{9728}{275}\right) p_\infty^2+\left(\frac{1762784}{11025}\log (2p_\infty )+\frac{224571 \pi ^2}{2240}-\frac{12181731011}{16464000}\right)p_\infty^3+O(p_\infty^4)$\\
%%%%%%%%%%%%%%%%%%%%%%%%%%%%%%%%%
$f_{\check u_2}^{G^5\, {\rm 1SF}}$&$\pi\left[-\frac{6}{p_\infty^6}+\frac{45}{p_\infty^4}+\frac{122}{5 p_\infty^3}+\frac{717}{2p_\infty^2}+\frac{13831}{280 p_\infty}+\frac{297 \pi^2}{20}+\frac{20457}{32}-\frac{64579}{5040}p_\infty+\left(\frac{335931}{560}-\frac{24993 \pi ^2}{1120}\right) p_\infty^2\right.$\\
&$\left.+\left(-\frac{10593}{350}  \log (\frac{p_\infty}{2} )+\frac{99 \pi ^2}{10}+\frac{29573617463 }{310464000}\right) p_\infty^3 + \left(
+\frac{1577 \pi ^2}{1120}-\frac{384 }{7}\right) p_\infty^4+O(p_\infty^5)\right]$\\
%%%%%%%%%%%%%%%
$f_{\check u_1}^{G^5\, {\rm 1SF}}$&$ \pi\left[ \frac{18}{p_\infty^6}-\frac{114}{p_\infty^4}-\frac{56}{5 p_\infty^3}
+\left(-\frac{3409}{4}-\frac{123 \pi^2}{64}\right)\frac{1}{p_\infty^2}
-\frac{15007}{210} p_\infty
+\frac{148}{5}\log (\frac{p_\infty}{2} )-\frac{49345 \pi ^2}{4096}-\frac{115753 }{96} 
-\frac{235867 p_\infty}{1260}\right.$\\
&$\left.+\left(\frac{5501}{70}  \log (\frac{p_\infty}{2} )-\frac{227435 \pi ^2}{8192}-\frac{884909}{4800} \right)p_\infty^2
-\frac{11646937}{63360}p_\infty^3
+\left(\frac{60521}{840}  \log (\frac{p_\infty}{2} )-\frac{219729011 \pi ^2}{8388608}+\frac{10171457  }{29400}\right)p_\infty^4 
+O(p_\infty^5)\right]$\\
\hline
$f_{b}^{G^5\, \overline {\rm 1SF}}$&$ -\frac{8}{p_\infty^9}+\frac{44}{p_\infty^7}+\frac{329+36 \pi ^2}{p_\infty^5}
+\frac{416}{45p_\infty^4}+\left(\frac{889}{6}+\frac{643 \pi^2}{4}\right)\frac{1}{p_\infty^3}+\left(\frac{169664}{1575}-\frac{47 \pi ^2}{5}\right)\frac{1}{p_\infty^2}$\\
&$+\left(\frac{12544 \log (2p_\infty)}{45}+\frac{14695 \pi ^2}{72}+\frac{575123}{2160}\right)\frac{1}{p_\infty}-\frac{1243 \pi ^2}{56}+\frac{159232}{3675}+\left(\frac{148864 \log (2p_\infty)}{525}-\frac{33019 \pi ^2}{480}+\frac{500415991}{252000}\right)p_\infty$\\
&$+\left(\frac{241 \pi^2}{120}-\frac{22294592}{363825}\right) p_\infty^2+\left(\frac{1762784 \log (2p_\infty)}{11025}+\frac{224571 \pi ^2}{2240}-\frac{33643816009}{49392000}\right)p_\infty^3$\\
%%%%
$f_{\check u_2}^{G^5\, \overline {\rm 1SF}}$& $\pi\left[-\frac{18}{p_\infty^6}+\frac{114}{p_\infty^4}+\frac{178}{5p_\infty^3}+\frac{\frac{3409}{4}+\frac{123 \pi ^2}{64}}{p_\infty^2}
+\frac{13847 }{120 p_\infty}-\frac{148}{5} \log (\frac{p_\infty}{2} ) +\frac{550853 \pi ^2}{20480}+\frac{115753}{96}+\frac{927433}{5040} p_\infty\right.$\\
&$+\left(-\frac{5501}{70}  \log (\frac{p_\infty}{2} )+\frac{85783 \pi ^2}{40960}+\frac{15166967}{33600}\right)p_\infty^2+\left(-\frac{10593}{350}  \log (\frac{p_\infty}{2} )+\frac{99 \pi^2}{10}+\frac{10356321807 }{34496000} \right)p_\infty^3$\\
&$\left.+ \left(-\frac{60521}{840}   \log (\frac{p_\infty}{2} )+\frac{1320222271 \pi ^2}{41943040}-\frac{50406217}{117600} \right)p_\infty^4+ O(p_\infty^5)\right]$\\
%%%%%
$f_{\check u_1}^{G^5\, \overline {\rm 1SF}}$&$\pi\left[  \frac{6}{p_\infty^6} - \frac{45}{p_\infty^4} -  
 \frac{717}{2 p_\infty^2}- \frac{20457}{32}    -  \frac{5385}{16}p_\infty^2 \right]$\\
\hline
$f_{b}^{G^5\,  {\rm 0SF}}$&$ -\frac{2}{p_\infty^9}+\frac{12}{p_\infty^7}+\frac{80+9 \pi ^2}{p_\infty^5}+\frac{\frac{81 \pi ^2}{2}-128}{p_\infty^3}+\left(\frac{225 \pi ^2}{8}-384\right)
  p_\infty+\frac{\frac{945 \pi ^2}{16}-576}{p_\infty}$\\
$f_{\check u_2}^{G^5\, {\rm 0SF}}$ & $0$\\ 
$f_{\check u_1}^{G^5\, {\rm 0SF}}$ & $ \frac{6 \pi }{p_\infty^6}-\frac{45 \pi }{p_\infty^4}-\frac{5385 \pi p_\infty^2}{16}-\frac{717 \pi }{2p_\infty^2}-\frac{20457 \pi }{32}$\\
\hline
$f_{b}^{G^5\, \overline {\rm 0SF}}$&$ -\frac{2}{p_\infty^9}+\frac{12}{p_\infty^7}+\frac{80+9 \pi ^2}{p_\infty^5}+\frac{\frac{81 \pi ^2}{2}-128}{p_\infty^3}+\left(\frac{225 \pi ^2}{8}-384\right)
   p_\infty+\frac{\frac{945 \pi ^2}{16}-576}{p_\infty}$\\
$f_{\check u_2}^{G^5\, \overline {\rm 0SF}}$&$ -\frac{6 \pi }{p_\infty^6}+\frac{45 \pi }{p_\infty^4}+\frac{5385 \pi p_\infty^2}{16}+\frac{717 \pi }{2 p_\infty^2}+\frac{20457 \pi }{32}$\\
$f_{\check u_1}^{G^5\, \overline {\rm 0SF}}$&  $0$\\
\end{tabular}
\end{ruledtabular}
\end{table*}

\section{Relations among $\hat E_5$ and $\hat P^{5,0}_{1\pm 2}$, $\hat P^{5,1}_{1\pm 2}$}

We have

\bea
\hat P^{5,0}_{1+2}&=& \frac{\hat E_5^{\rm 0SF}}{(\gamma+1)(\gamma^2-1)^{5/2}}-\frac{\gamma-1}{\gamma+1}\hat P^5_{1-2}\,,\nonumber\\
\hat P^{5,1}_{1+2}&=& \frac{\hat E_5^{\rm 1SF}}{(\gamma+1)(\gamma^2-1)^{5/2}}+4\frac{\gamma-1}{\gamma+1}\hat P^5_{1-2}\,,
\eea
so that, for example,
\beq
4\hat P^{5,0}_{1+2}+\hat P^{5,1}_{1+2}=\frac{4\hat E_5^{\rm 0SF}+\hat E_5^{\rm 1SF}}{(\gamma+1)(\gamma^2-1)^{5/2}}\,.
\eeq

\end{document}